\documentclass[12pt, a4paper]{article}

\usepackage{amssymb}
\usepackage{amsmath}

\font\small=cmr10

\newcommand{\qed}{\hfill $\square$}
\newcommand\unit{\hbox{\rm 1\kern-2.8truept l}}
\newcommand\re{{\Re}\kern-1pt e}
\newcommand\im{{\Im}\kern-1pt m}
\newcommand\tr{\hbox{\rm tr}}
\newcommand{\Bh}{\mathcal{B}(\h) }

\newcommand{\pL}{\mathcal{L}^\prime}
\newcommand{\pT}{\mathcal{T}^\prime}

\newcommand{\trhi}[1]{\theta  #1\theta}

\newcommand\Lform{{\mathcal{L}}\kern-7.56pt\raise1.0pt\hbox{$-$}}

\newcommand{\h}{\mathsf{h}}
\newcommand{\T}{\mathcal{T}}
\newcommand{\Ll}{\mathcal{L}}
\newcommand{\radp}{\sqrt{1-\nu}+\sqrt{\nu}}
\newcommand{\radm}{\sqrt{1-\nu}-\sqrt{\nu}}
\newcommand{\ketbra}[2]{\mid\! #1\rangle\langle #2\!\mid}
\newtheorem{definition}{Definition}
\newtheorem{theorem}[definition]{Theorem}
\newtheorem{proposition}[definition]{Proposition}

\newtheorem{remark}{Remark}
\newtheorem{lemma}[definition]{Lemma}
\newtheorem{corollary}[definition]{Corollary}

\begin{document}

\title{Generators of KMS Symmetric Markov Semigroups 
on $\Bh$ Symmetry and Quantum Detailed Balance}
\date{}
%\author{F. Fagnola and V. Umanit\`a}

\maketitle

\centerline{\small{F. FAGNOLA}} 
\centerline{\small{Dipartimento di Matematica, Politecnico di Milano,}} 
\centerline{\small{Piazza Leonardo da Vinci 32, I-20133 Milano (Italy)}}
\centerline{\small{franco.fagnola@polimi.it}}
\medskip
\centerline{\small{V. UMANIT\`A}} 
\centerline{\small{Dipartimento di Matematica, Universit\`a di Genova,}} 
\centerline{\small{Via Dodecaneso 35, I-16146 Genova (Italy)}}
\centerline{\small{veronica.umanita@fastwebnet.it}}

\begin{abstract}
We find the structure of generators of norm continuous 
quantum Markov semigroups on $\Bh$ that are symmetric with 
respect to the scalar product $\tr(\rho^{1/2}x^*\rho^{1/2}y)$ 
induced by a faithful normal invariant state invariant state 
$\rho$ and satisfy two quantum generalisations of the 
classical detailed balance condition related with this 
non-commutative notion of symmetry: the so-called 
standard detailed balance condition and the standard 
detailed balance condition with an antiunitary time reversal. 
\end{abstract}

\section{Introduction}
Symmetric Markov semigroups have been extensively studied in 
classical stochastic analysis (Fukushima et al. \cite{FuOsTa} 
and the references therein) because their generators and 
associated Dirichlet forms are very well tractable by 
Hilbert space and probabilistic methods. 

Their non-commutative counterpart has also been deeply 
investigated (Albeverio and Goswami \cite{AlbGos}, 
Cipriani \cite{Cip}, Davies and Lindsay \cite{DaLi}, 
Goldstein and Lindsay \cite{GoLi95}, Guido, Isola and 
Scarlatti \cite{GIS}, Park \cite{Park}, Sauvageot \cite{Sa} 
and the references therein). 

The classical notion of symmetry with respect to a measure, 
however, admits several non-commutative generalisations. 
Here we shall consider the so-called KMS-symmetry that seems  
more natural from a mathematical point of view (see e.g. 
Accardi and Mohari \cite{AcMo}, 
Cipriani \cite{Cip}, \cite{Cip2}, Goldstein and Lindsay \cite{GoLi93}, 
Petz \cite{Petz}) and find the structure of generators 
of norm-continuous quantum Markov semigroups (QMS) 
on the von Neumann algebra $\Bh$ of all bounded operators 
on a complex separable Hilbert space $\h$ that are symmetric 
or satisfy quantum detailed balance conditions associated with 
KMS-symmetry or generalising it. 

We consider QMS on $\Bh$, i.e. weak$^*$-continuous semigroups 
of normal, completely positive, identity preserving maps 
$\T=(\T_t)_{t\ge 0}$ on $\Bh$, with a faithful normal 
invariant state $\rho$. This defines pre-scalar products 
on $\Bh$ by $(x,y)_s=\tr(\rho^{1-s} x^*\rho^{s}y)$ for 
$s\in[0,1]$ and allows one to define the $s$-dual semigroup 
$\T^\prime$ on $\Bh$ satisfying $\tr(\rho^{1-s} x^*\rho^{s}\T_t(y))
=\tr(\rho^{1-s} \T^\prime_t(x)^*\rho^{s}y)$ for 
all $x,y\in\Bh$. The above scalar products coincide 
on an Abelian von Neumann algebra, the notion 
of symmetry $\T=\T^\prime$, however, clearly depends on 
the choice of the parameter $s$. 

The most studied cases are $s=0$ and $s=1/2$. Denoting 
$\T_*$ the predual semigroup, a simple computation 
yields $\T^\prime(x)= \rho^{-(1-s)} 
\T_{*t}(\rho^{1-s} x\rho^s)\rho^{-s}$, and shows 
that for $s=1/2$ the maps $\T^\prime_t$ are positive but, 
for $s\not=1/2$ this may not be the case. Indeed, it is 
well-known that, for $s\not=1/2$, the maps $\T^\prime_t$ 
are positive if and only if the maps $\T_t$ commute with 
the modular group $(\sigma_t)_{t\in\mathbb{R}}$, 
$\sigma_t(x)=\rho^{it}x\rho^{-it}$ (see e.g. \cite{KFGV} 
Prop. 2.1 p. 98, \cite{MaSt} Th. 6 p. 7985, for $s=0$,   
\cite{FFVU} Th. 3.1 p. 341, Prop. 8.1 p. 362 for $s\not=1/2$).
This quite restrictive condition implies that the generator 
has a very special form that makes simpler the mathematical 
study of symmetry but imposes strong structural constraints 
(see e.g. \cite{KFGV} and \cite{FFVU-Bedlewo08}).

Here we shall consider the most natural choice $s=1/2$ 
whose consequences are not so stringent and say that
$\T$ is {\em KMS-symmetric} if it coincides with its 
dual $\T^\prime$. KMS-symmetric QMS were introduced by 
Cipriani \cite{Cip} and Goldstein and Lindsay \cite{GoLi93}; 
we refer to \cite{Cip2} for a discussion of the connection with 
the KMS condition justifying this terminology. 

All quantum versions of the classical principle of detailed 
balance (Agarwal \cite{Agarwal}, Alicki \cite{alicki}, Frigerio, Gorini, 
Kossakowski and Verri \cite{KFGV}, Majewski \cite{Maje}, \cite{Maje2}), 
which is at the basis of equilibrium physics, are formulated prescribing 
a certain relationship between $\T$ and $\T^\prime$ or between their generators, 
therefore they depending of the underlying notion of symmetry.
This work clarifies the structure of generators of QMS that are 
KMS symmetric or satisfy a quantum detailed balance condition 
involving the above scalar product with $s=1/2$ and is a key step 
towards understanding which is the most natural and flexible in view 
of the study of their generalisations for quantum systems out of 
equilibrium as, for instance, the {\em dynamical} detailed balance 
condition introduced by Accardi and Imafuku \cite{AcIm}.

The generator $\Ll$ of a norm-continuous QMS can 
be written in the standard Gorini-Kossakowski-Sudarshan 
\cite{GoKoSu} and Lindblad \cite{Lindblad} (GKSL) form 
\begin{equation}\label{eq-GKSL}
\Ll(x)=i[H,x] - \frac{1}{2}
\sum_{\ell\ge 1}
\left(L^*_\ell L_\ell x-2L^*_\ell xL_\ell + xL^*_\ell L_\ell\right)
\end{equation}
where $H, L_\ell\in\Bh$ with $H=H^*$ and the series 
$\sum_{\ell\ge 1} L_\ell^*L_\ell$ is strongly convergent. 
The operators $L_\ell, H$ in (\ref{eq-GKSL}) are not uniquely 
determined by $\Ll$, however, under a natural minimality 
condition (Theorem \ref{th-special-GKSL} below) and a zero-mean 
condition $\tr(\rho L_\ell)=0$ for all $\ell\ge 1$, $H$ is 
determined up to a scalar multiple of the identity operator and 
the $(L_\ell)_{\ell\ge 1}$ up to a unitary transformation of 
the multiplicity space of the completely positive part of $\Ll$. 
We shall call {\em special} a GKSL representation of $\Ll$ 
by operators $H,L_\ell$ satisfying these conditions.

As a result, by the remark following Theorem \ref{th-special-GKSL}, 
in a special GKSL representation 
of $\Ll$, the operator $G=-2^{-1}\sum_{\ell\ge 1}L^*_\ell L_\ell-iH$, 
is uniquely determined by $\Ll$ up to a purely imaginary multiple 
of the identity operator and allows us to write $\Ll$ in the form 
\begin{equation}\label{eq-GKSL-G} 
\Ll(x) = G^* x + \sum_{\ell\ge 1} L^*_\ell x L_\ell + x G. 
\end{equation} 

Our characterisation of QMS that are KMS-symmetric or  
satisfy a quantum detailed balance condition generalising 
related with KMS-symmetry is given in terms of the operators 
$G,L_\ell$ (or, in an equivalent way $H,L_\ell$) of a special 
GKSL representation. 

Theorem \ref{th-KMSsimm} shows that a QMS  is KMS-symmetric 
if and only if the operators $G,L_\ell$ of a special GKSL 
representation of its generator satisfy $\rho^{1/2}G^* = G\rho^{1/2} 
+ic\rho^{1/2}$ for some $c\in\mathbb{R}$ and $\rho^{1/2}L^*_k 
=\sum_\ell u_{k\ell}L_\ell\rho^{1/2}$ for all $k$ and some 
unitary $(u_{k\ell})$ on the multiplicity space of the completely 
positive part of $\Ll$ coinciding with its transpose, 
i.e. such that $u_{k\ell}=u_{\ell k}$ for all $k,\ell$. 

In order to describe our results on the structure of generators 
of QMS satisfying a quantum detailed balance condition we first 
recall some basic definitions. The best known is due 
to Alicki \cite{alicki} and Frigerio-Gorini-Kossakowski-Verri 
\cite{KFGV}: a norm-continuous QMS $\T=(\T_t)_{t\ge 0}$ on $\Bh$ 
satisfies the {\em Quantum Detailed Balance} (QDB) condition  
if there exists an operator $\widetilde{\Ll}$ on $\Bh$ and 
self-adjoint operator $K$ on $\h$ such that $\tr(\rho\widetilde{\Ll}(x)y)
=\tr(\rho x \Ll(y))$ and $\Ll(x)-\widetilde{\Ll}(x)=2i[K,x]$ for all 
$x,y\in\Bh$. 
Roughly speaking we can say that $\Ll$ satisfies the QDB condition 
if the difference of $\Ll$ and its adjoint $\widetilde{\Ll}$ with 
respect to the pre-scalar product on $\Bh$ given by $\tr(\rho a^*b)$ 
is a derivation.

This QDB implies that the operator $\widetilde{\Ll}=\Ll-2i[K,\cdot\,]$ 
is conditionally completely positive and then generates a QMS 
$\widetilde{\T}$. Therefore $\Ll$ and the maps $\T_t$ commute 
with the modular group. This restriction does not follow if the 
dual QMS is defined with respect to the symmetric pre-scalar product 
with $s=1/2$. 

The QDB can be readily reformulated replacing $\widetilde{\Ll}$ 
with the adjoint $\Ll^\prime$ defined via the symmetric 
scalar product; the resulting condition will be called 
{\em Standard Quantum Detailed Balance} condition (SQDB) 
(see e.g. \cite{DeFr}). 

Theorem \ref{th-SQDB} characterises generators $\Ll$ satisfying 
the SQDB and extends previous partial results by Park \cite{Park} 
and the authors \cite{FFVU}: the SQDB holds if and only if there 
exists a unitary matrix $(u_{k\ell})$, coinciding with its transpose, 
i.e. $u_{k\ell}=u_{\ell k}$ for all $k,\ell$, such that $\rho^{1/2}L^*_k 
=\sum_{\ell}u_{k\ell}L_\ell\rho^{1/2}$. This shows, in particular, 
that the SQDB depends only on the $L_\ell$'s and does not involve 
directly $H$ and $G$. Moreover, we find explicitly the unitary 
$(u_{k\ell})_{k\ell}$ providing also a geometrical characterisation of 
the SQDB (Theorem \ref{th-SQDB-geo}) in terms of the operators 
$L_\ell\rho^{1/2}$ and their adjoints as Hilbert-Schmidt operators 
on $\mathsf{h}$.

We also consider (Definition \ref{def-SQDB-theta}) another notion 
of quantum detailed balance, inspired by the original Agarwal's 
notion (see \cite{Agarwal}, Majewski \cite{Maje}, \cite{Maje2}, 
Talkner \cite{Talk}) involving an antiunitary {\em time reversal} 
operator $\theta$ which does not play any role in Alicki et al. 
definition. Time reversal appears to keep into the account the parity 
of quantum observables; position and energy, for instance, are even, 
i.e. invariant under time reversal, momentum are odd, i.e. change 
sign under time reversal. The original Agarwal's definition, however, 
depends on the $s=0$ pre-scalar product and implies then, that a 
QMS satisfying this quantum detailed balance condition must commute 
with the modular automorphism. Here we study the modified version 
(Definition \ref{def-SQDB-theta}) involving the symmetric $s=1/2$ 
pre-scalar product that we call the SQDB-$\theta$ condition.

Theorem \ref{th-SQDB-TR} shows that $\Ll$ satisfies the 
SQDB-$\theta$ condition if and only if there exists a special 
GKSL representation of $\Ll$ by means of operators $H,L_\ell$ 
such that $G\rho^{1/2}=\rho^{1/2}\theta{G^*}\theta$ and a 
unitary self-adjoint $(u_{k\ell})_{k\ell}$ such that 
$\rho^{1/2}{L}_k^* = \sum_\ell u_{k\ell}\theta{L_\ell}\theta\rho^{1/2}$ 
for all $k$. Here again $(u_{k\ell})_{k\ell}$ is explicitly determined  
by the operators $L_\ell\rho^{1/2}$ (Theorem \ref{th-SQDB-theta-geo}). 

We think that these results show that the SQDB condition is 
somewhat weaker than the SQDB-$\theta$ condition because  
the first does not involve the directly the operators $H$, $G$. 
Moreover, the unitary operator in the linear relationship 
between $L_\ell\rho^{1/2}$ and their adjoints is transpose 
symmetric and any point of the unit disk could be in its spectrum 
is while, for generators satisfying the SQDB-$\theta$, it is 
self-adjoint and its spectrum is contained in $\{-1,1\}$. 
Therefore, by the spectral theorem, it is possible in principle 
to find a standard form for the generators of QMSs satisfying the 
SQDB-$\theta$ generalising the standard form of generators satisfying 
the usual QDB condition (that commute with the modular group) 
as illustrated in the case of QMSs on $M_2({\mathbb{C}})$ 
studied in the last section. This classification must be much 
more complex for generators of QMSs satisfying the SQDB.

The above arguments and the fact that the SQDB-$\theta$ condition 
can be formulated in a simple way both on the QMS or on its generator 
(this is not the case for the QDB-$\theta$ when $\Ll$ and its 
Hamiltonian part $i[H,\cdot]$ do not commute), lead us to the 
conclusion that the SQDB-$\theta$ is the more natural non-commutative 
version of the classical detailed balance condition. 

The paper is organised as follows. In Section \ref{sect-dualQMS-SQDB} 
we construct the dual QMS $\T^\prime$ and recall the quantum detailed 
balance conditions we investigate, then we study the relationship between 
the generators of a QMS and its adjoint in Section \ref{sect-genQMS-and-dual}. 
Our main results on the structure of generators are proved in 
Sections \ref{sect-SDB-QMS} (QDB without time reversal) and 
\ref{sect-SDB-TR-QMS} (with time reversal).

\section{The dual QMS, KMS-symmetry and quantum detailed balance}
\label{sect-dualQMS-SQDB}

We start this section by constructing the dual semigroup of a 
norm-continuous QMS with respect to the $(\cdot,\cdot)_{1/2}$ 
pre-scalar product on $\Bh$ defined by an invariant state 
$\rho$ and prove some properties that will be useful in 
the sequel. Although this result may be known, the presentation 
given here leads in a simple and direct way to the dual 
QMS avoiding non-commutative $L^p$-spaces techniques. 

\begin{proposition}\label{prop-dual-map}
Let $\varPhi$ be a positive unital normal map on $\Bh$ with a 
faithful normal invariant state $\rho$. There exists a unique 
positive unital normal map $\varPhi^\prime$ on $\Bh$ such that 
\[
\tr\left(\rho^{1/2}\varPhi^\prime(x)\rho^{1/2} y\right)
=\tr\left(\rho^{1/2}x\rho^{1/2} \varPhi(y)\right)
\]
for all $x,y\in\Bh$. If $\varPhi$ is completely positive, 
then $\varPhi^\prime$ is also completely positive.
\end{proposition}

\noindent{\bf Proof.} Let $\varPhi_*$ be the predual map 
on the Banach space of trace class operators on $\h$ and
let $Rk(\rho^{1/2})$ denote the range of the operator 
$\rho^{1/2}$. This is clearly dense in $\h$ because $\rho$ 
is faithful and coincides with the domain of the unbounded 
self-adjoint operator $\rho^{-1/2}$. 

For all self-adjoint $x\in\Bh$ consider the sesquilinear 
form on the domain $Rk(\rho^{1/2})\times Rk(\rho^{1/2})$
\[
F(v,u) = \langle \rho^{-1/2}v,
\varPhi_*(\rho^{1/2}x\rho^{1/2})\rho^{-1/2}u\rangle. 
\]
By the invariance of $\rho$ and positivity of $\varPhi_{*}$ 
we have 
\[
-\Vert x \Vert \rho 
= - \Vert x \Vert \varPhi_*(\rho) 
\le\varPhi_*(\rho^{1/2}x\rho^{1/2})
\le \Vert x \Vert \varPhi_*(\rho) 
=\Vert x \Vert \rho.
\]
Therefore $|F(u,u)|\le \Vert x \Vert \cdot\Vert v\Vert \cdot \Vert u\Vert$. 
Thus sesquilinear form is bounded and there exists a unique bounded 
operator $y$ such that, for all $u,v\in Rk(\rho^{1/2})$, 
\[
\langle v, y u \rangle = \langle \rho^{-1/2}v,\varPhi_*(\rho^{1/2}x\rho^{1/2})\rho^{-1/2}u\rangle.
\]
Note that, $\varPhi$ being a ${}^*$-map, and $x$ self/adjoint
\begin{eqnarray*}
\langle v, y^* u \rangle & = & \overline{\langle y^*u,v\rangle} \\
& = & \overline{\langle \rho^{-1/2}u,\varPhi_{*}(\rho^{1/2}x\rho^{1/2})\rho^{-1/2}v\rangle}\\
& = & \overline{\langle \varPhi_{*}(\rho^{1/2}x\rho^{1/2})\rho^{-1/2}u,\rho^{-1/2}v\rangle}\\
& = & \langle \rho^{-1/2}v,\varPhi_{*}(\rho^{1/2}x\rho^{1/2})\rho^{-1/2}u\rangle.
\end{eqnarray*}
This shows that $y$ is self-adjoint. Defining $\varPhi^\prime(x):=y$, 
we find a real-linear map on self-adjoint operators on $\Bh$ that can 
be extended to a linear map on $\Bh$ decomposing each 
self-adjoint operator as the sum of its self-adjoint and anti self-adjoint 
parts.

Clearly $\varPhi^\prime$ is positive because 
$\rho^{1/2}\varPhi^\prime(x^*x)\rho^{1/2} 
= \varPhi_{*}(\rho^{1/2}x^*x\rho^{1/2})$ 
and $\varPhi_{*}$ is positive. 
Moreover, by the above construction $\varPhi^\prime(\unit)=\unit$, i.e. 
$\varPhi^\prime$ is unital. 
Therefore $\Phi^\prime$ is a norm-one contraction.

If $\varPhi$ is completely positive, then $\varPhi_{*}$ is also and formula 
$\rho^{1/2}\varPhi^\prime(x)\rho^{1/2}= \varPhi_{*}(\rho^{1/2}x\rho^{1/2})$ 
shows that $\varPhi^\prime$ is completely positive.

Finally we show that $\varPhi^\prime$ is normal. Let 
$(x_\alpha)_{\alpha}$ be a net of positive operators on 
$\Bh$ with least upper bound $x\in\Bh$. For all $u\in \h$ 
we have then 
\begin{eqnarray*}
\sup_\alpha \langle \rho^{1/2}u,\varPhi^\prime(x_\alpha)\rho^{1/2}u\rangle 
& = & \sup_\alpha \langle u,\varPhi_{*}(\rho^{1/2}x_\alpha\rho^{1/2})u\rangle \\
& = & \langle u,\varPhi_{*}(\rho^{1/2}x\rho^{1/2})u\rangle 
= \langle \rho^{1/2}u,\varPhi^\prime(x)\rho^{1/2}u\rangle.
\end{eqnarray*}
Now if $u\in\h$, for every $\varepsilon>0$, we can find a 
$u_\varepsilon\in Rk(\rho^{1/2})$ such that 
$\Vert u - u_\varepsilon\Vert < \varepsilon$ by the density 
of the range of $\rho^{1/2}$. We have then
\begin{eqnarray*}
\left|\langle u, 
  \left(\varPhi^\prime(x_\alpha)-\varPhi^\prime(x)\right)u\rangle\right|
& \le & \varepsilon \left\Vert\,\varPhi^\prime(x_\alpha)
-\varPhi^\prime(x)\right\Vert 
               \left(\Vert u\Vert + \Vert u_\varepsilon\Vert\right)\\
& + & \left|\langle u_\varepsilon, 
        \left(\varPhi^\prime(x_\alpha)
             -\varPhi^\prime(x)\right)u_\varepsilon\rangle\right|
\end{eqnarray*} 
for all $\alpha$. The conclusion follows from the arbitrarity of 
$\varepsilon$ and the uniform boundedness of 
$\Vert\,\varPhi^\prime(x_\alpha)-\varPhi^\prime(x)\Vert$ 
and $\Vert u_\varepsilon\Vert$. \quad \qed

\begin{theorem}\label{th-dual-QMS}
Let $\T$ be a QMS on $\Bh$ with a faithful normal invariant 
state $\rho$. There exists a QMS $\T^\prime$ on $\Bh$ such that 
\begin{equation}\label{eq-duality-T}
\rho^{1/2}\T_t^\prime(x)\rho^{1/2}=\T_{*t}(\rho^{1/2} x \rho^{1/2})
\end{equation}
for all $x\in\Bh$ and all $t\ge 0$. 
\end{theorem}

\noindent{\bf Proof.} By Proposition \ref{prop-dual-map}, for 
each $t\ge 0$, there exists a unique completely positive normal 
and unital contraction $\T^\prime_t$ on $\Bh$ satisfying (\ref{eq-duality-T}). 
The semigroup property follows form the algebraic computation 
\begin{eqnarray*}
\rho^{1/2}\T^\prime_{t+s}(x)\rho^{1/2} & = & 
\T_{*t}\left(\T_{*s}(\rho^{1/2}x\rho^{1/2})\right) \\
& = & \T_{*t}\left(\rho^{1/2}\T^\prime_s(x)\rho^{1/2})\right)= 
\rho^{1/2}\T^\prime_{t}\left(\T^\prime_s(x))\right)\rho^{1/2}. 
\end{eqnarray*}
Since the map $t\to \langle \rho^{1/2}v,\T^\prime_t(x)\rho^{1/2}u\rangle$ 
is continuous by the identity (\ref{eq-duality-T}) for all $u,v\in\h$, and 
$\Vert\T^\prime_t(x)\Vert\le \Vert x \Vert$ for all $t\ge 0$, a 
$2\varepsilon$ approximation argument shows that $t\to \T^\prime_t(x)$ 
is continuous for the weak$^*$-operator topology on $\Bh$. It follows that 
$\T^\prime=(\T^\prime_t)_{t\ge 0}$ is a QMS on $\Bh$. \qed

\begin{definition}\label{def-symmetric-dual} 
The quantum Markov semigroup $\T^\prime$ is called the \emph{dual semigroup} 
of $\T$ with respect to the invariant state $\rho$. 
\end{definition}

It is easy to see, using (\ref{eq-duality-T}), that 
$\rho$ is an invariant state also for $\T^\prime$. 

\begin{remark} {\rm When $\T$ is norm-continuous it is 
not clear whether also $\T^\prime$ is norm-continuous. 
Here, however, we are interested in generators of symmetric 
or detailed balance QMS. We shall see that these additional 
properties of $\T$ imply that also $\T^\prime$ is norm continuous. 
Therefore we proceed studying norm-continuous QMSs whose dual 
is also norm-continuous.}
\end{remark}

The quantum detailed balance condition of Alicki, Frigerio, Gorini, 
Kossakowski and Verri modified by considering the pre-scalar 
product $(\cdot,\cdot)_{1/2}$ on $\Bh$, usually called {\em standard} 
(see e.g. \cite{DeFr}) because of multiplications by $\rho^{1/2}$ 
as in the standard representation of $\Bh$, is defined as follows. 

\begin{definition}\label{def-SQDB}
The QMS $\mathcal{\T}$ generated by $\Ll$ satisfies the 
{\em standard quantum detailed balance condition} (SQDB) 
if there exists an operator $\Ll^\prime$ on $\Bh$ and a 
self-adjoint operator $K$ on $\h$ such that
\begin{equation}\label{eq-SQDB}
\tr(\rho^{1/2}x\rho^{1/2} \Ll(y))=
\tr(\rho^{1/2}\Ll^\prime(x)\rho^{1/2} y), \qquad
\Ll(x) - \Ll^\prime(x)= 2i[K,x]
\end{equation} 
for all $x\in\Bh$.  
\end{definition}

The operator $\Ll^\prime$ in the above definition must be norm-bounded 
because it is everywhere defined and norm closed. To see this consider 
a sequence $(x_n)_{n\ge 1}$ in $\Bh$ converging in norm to a $x\in \Bh$ 
such that $(\Ll(x_n))_{n\ge 1}$ converges in norm to $b\in \Bh$ and note 
that 
\begin{eqnarray*} 
\tr\left(\rho^{1/2}\Ll^\prime(x)\rho^{1/2} y\right) 
& = & \lim_{n\to\infty}\tr\left(\rho^{1/2}x_n\rho^{1/2}\Ll(y)\right) \\ 
& = & \lim_{n\to\infty}\tr\left(\rho^{1/2}\Ll^\prime(x_n)\rho^{1/2}y\right) 
= \tr\left(\rho^{1/2}b\rho^{1/2}y\right) 
\end{eqnarray*} 
for all $y\in\Bh$. The elements $\rho^{1/2}y\rho^{1/2}$, with $y\in\Bh$, 
are dense in the Banach space of trace class operators on $\h$ because 
$\rho$ is faithful. Therefore shows that $\Ll^\prime(x)=b$ and $\Ll^\prime$ 
is closed. 

Since both $\Ll$ and $\Ll^\prime$ are bounded, also $K$ is bounded. 

\medskip
We now introduce another definition of quantum detailed balance, 
due to Agarwal \cite{Agarwal} with the $s=0$ pre-scalar product, 
that involves a {\em time reversal} $\theta$. This is an antiunitary 
operator on $\h$, i.e. $\langle\theta u,\theta v\rangle = 
\langle v,u\rangle$ for all $u,v\in\mathsf{h}$,  
such that $\theta^2=\unit$ and $\theta^{-1} =\theta^*=\theta$. 

Recall that, $\theta$ is antilinear, i.e. $\theta z u = \bar{z}u$ 
for all $u\in\mathsf{h}$, $z\in{\mathbb{C}}$, and its adjoint 
$\theta^*$ satisfies $\langle u,\theta v\rangle=\langle v, 
\theta^*u\rangle$ for all $u,v\in\h$. Moreover $\theta\, x\,\theta $ 
belongs to $\Bh$ (linearity is re-established) and 
$\tr(\theta\, x\theta )=\tr(x^*)$ for every trace-class 
operator $x$ (\cite{FFVUTR} Prop. 4), indeed, taking 
an orthonormal basis of $\h$, we have 
\begin{eqnarray*}
\tr(\theta x\theta )=\sum_j\langle e_j,\theta x\theta e_j\rangle
& = & \sum_j\langle x\theta e_j,\theta^*e_j \rangle \\ 
& = &\sum_j\langle \theta e_j,x^*\theta^*e_j \rangle=\tr(x^*).
\end{eqnarray*}
It is worth noticing that the cyclic property of the trace 
does not hold for $\theta$, since $\tr(\theta\, x\theta )=\tr(x^*)$ 
may not be equal to $\tr(x)$ for non 
self-adjoint $x$.

\begin{definition}\label{def-SQDB-theta} 
The QMS $\mathcal{\T}$ generated by $\Ll$ satisfies the standard 
quantum detailed balance condition with respect to the time reversal 
$\theta$ (SQDB-$\theta$) if 
\begin{equation}\label{eq-SQDB-theta} 
\tr(\rho^{1/2}x\rho^{1/2} \Ll(y))= 
\tr(\rho^{1/2}\theta y^* \theta \rho^{1/2} 
\Ll(\theta x^*\theta )), 
\end{equation} 
for all $x,y\in\Bh$. 
\end{definition} 

The operator $\theta$ is used to keep into the account parity 
of the observables under time reversal. Indeed, a self-adjoint 
operator $x\in\Bh$ is called \emph{even} (resp. \emph{odd}) if 
$\theta x\theta =x$ (resp. $\theta x\theta =-x$). The typical 
example of antilinear time reversal is a conjugation (with respect 
to some orthonormal basis).

This condition is usually stated (\cite{Maje}, \cite{Maje2}, 
\cite{Talk}) for the QMS $\T$ as 
\begin{equation}\label{eq-SQDB-theta-T}
\tr(\rho^{1/2}x\rho^{1/2} \T_t(y))= 
\tr(\rho^{1/2}\theta y^* \theta \rho^{1/2} 
\T_t(\theta x^*\theta )), 
\end{equation}
for all $t\ge 0$, $x,y\in\Bh$. In particular, for $t=0$ we find that 
this identity holds if and only if $\rho$ and $\theta$ commute, 
i.e. $\rho$ is an even observable. This is the case, for instance, 
when $\rho$ is a function of the energy.

\begin{lemma}\label{lem-rho-theta-comm} 
The following conditions are equivalent: 
\begin{enumerate}
\item[(i)] $\theta$ and $\rho$ commute, 
\item[(ii)] $\tr(\rho^{1/2}x\rho^{1/2} y) 
= \tr(\rho^{1/2}\theta y^* \theta \rho^{1/2} \theta x^*\theta )$ 
for all $x,y\in\Bh$.
\end{enumerate}
\end{lemma}

\noindent{\bf Proof.} If $\rho$ and $\theta$ commute, from 
$\tr(\theta a \theta )=\tr(a^*)$, we have  
\[
\tr(\rho^{1/2}\theta y^* \theta \rho^{1/2} \theta x^*\theta ) 
= \tr(\theta(\rho^{1/2}\theta y^* \rho^{1/2}x^*)\theta )
= \tr(x\rho^{1/2}y\rho^{1/2})
\]
and (ii) follows cycling $\rho^{1/2}$. 
Conversely, if (ii) holds, taking $x=\unit$, we have 
\[
\tr(\rho y)=\tr(\rho \theta y^*\theta )
=\tr\left(\theta (\theta y^*\theta )^*\rho\theta \right)
=\tr\left( y\theta\rho\theta \right)=\tr(\theta\rho\theta y),
\]
for all $y\in\Bh$, and $\rho= \theta\rho\theta $. \qed
\smallskip

\begin{proposition}\label{prop-SQDB-theta-L-T}
If $\rho$ and $\theta$ commute then (\ref{eq-SQDB-theta}) 
and (\ref{eq-SQDB-theta-T}) are equivalent. 
\end{proposition}

\noindent{\bf Proof.} Clearly (\ref{eq-SQDB-theta}) follows from 
(\ref{eq-SQDB-theta-T}) differentiating at $t=0$. 

Conversely, putting $\alpha(x)=\theta x \theta $ and denoting 
$\Ll_*$ the predual of $\Ll$ we can write (\ref{eq-SQDB-theta}) as 
\[
\tr(\Ll_*(\rho^{1/2}x\rho^{1/2})y)
=\tr\left(\rho^{1/2}\alpha(y^*)\rho^{1/2}\Ll(\alpha(x^*))\right)
=\tr\left( \rho^{1/2}\alpha(\Ll(\alpha(x)))\rho^{1/2}y\right),
\]
for all $y\in\Bh$, because $\tr(\alpha(a))=\tr(a^*)$. Therefore we have 
\[
\Ll_*(\rho^{1/2}x\rho^{1/2})=\rho^{1/2}\alpha(\Ll(\alpha(x)))\rho^{1/2}
\]
and, iterating, $\Ll_*^n(\rho^{1/2}x\rho^{1/2})= 
\rho^{1/2}\alpha(\Ll^n(\alpha(x)))\rho^{1/2}$ for all $n\ge 1$. 
It follows that (\ref{eq-SQDB-theta}) holds for all powers 
$\Ll^n$ with $n\ge 1$. Since $\rho$ and $\theta$ commute, it 
is true also for $n=0$ and we find (\ref{eq-SQDB-theta-T}) 
by the exponentiation formula $\T_t=\sum_{n\ge 0}t^n\Ll^n/n!$. \qed
\smallskip

We do not know whether the SQDB condition (\ref{eq-SQDB}) of 
Definition \ref{def-SQDB} has a simple explicit formulation in terms 
of the maps $\T_t$ if $\Ll$ and $\Ll^\prime$ do not commute. 

\begin{remark} {\rm 
The SQDB condition (\ref{eq-SQDB-theta}), by $\tr(\theta a\theta ) 
=\tr(a^*)$, reads 
\[
\tr(\rho^{1/2}x\rho^{1/2} \Ll(y))= 
\tr(\rho^{1/2}(\theta \Ll(\theta x\theta)\theta)\rho^{1/2} x), 
\]
for all $x,y\in\Bh$, i.e. $\Ll^\prime(x)=\theta \Ll(\theta x\theta)\theta$.

Write $\Ll$ in a special GKSL form as in (\ref{eq-GKSL}) and 
decompose the generator $\Ll=\Ll_0 + i[H,\cdot\,]$ into the sum of its 
dissipative part $\Ll_0$ and derivation part $i[H,\cdot\,]$. If $H$ 
commutes with $\theta$, by the antilinearity of $\theta$, we find 
$\Ll^\prime(x) = \theta \Ll_0(\theta x\theta)\theta - i [H,x]$. 
Therefore, if the dissipative part is time reversal invariant, i.e. 
$\Ll_0(x) = \theta \Ll_0(\theta x\theta)\theta$, we end up  
with $\Ll^\prime = \Ll - 2i[H,\cdot\,]$. 

The relationship with Definition \ref{def-SQDB} of SQDB, in this case, 
is then clear. 
The SQDB conditions of Definition \ref{def-SQDB} and \ref{def-SQDB-theta}, 
however, in general are not comparable.
}
\end{remark}

\section{The generator of a QMS and its dual}\label{sect-genQMS-and-dual}

We shall always consider {\em special}  GKSL representations of 
the generator of a norm-continuous QMS by means of operators $L_\ell,H$. 
These are described by the following theorem (we refer to \cite{Partha} 
Theorem 30.16 for the proof).

\begin{theorem}\label{th-special-GKSL} 
Let $\Ll$ be the generator of a norm-continuous QMS on 
$\Bh$ and let $\rho$ be a normal state on $\Bh$. There exists 
a bounded self-adjoint operator $H$ and a finite or infinite 
sequence $(L_\ell)_{\ell\ge 1}$ of elements of $\Bh$ such that: 
\begin{enumerate}
\item[{\rm (i)}] $\hbox{\rm tr}(\rho L_\ell)=0$ for each $\ell\geq 1$, 
\item[{\rm (ii)}] $\sum_{\ell\geq 1}L^*_\ell L_\ell$ is a strongly 
convergent sum,
\item[{\rm (iii)}] if $\sum_{\ell\geq 0}\vert c_\ell\vert^2<\infty$ 
and $c_0+\sum_{\ell \geq 1}c_\ell L_\ell=0$ 
for complex scalars $(c_k)_{k\geq 0}$ then $c_k=0$ for every $k\geq 0$,
\item[{\rm (iv)}] the GKSL representation (\ref{eq-GKSL}) holds.
\end{enumerate}
If $H^\prime,(L^\prime_\ell)_{\ell\ge 1}$ is another family of
bounded operators in $\Bh$ with $H^\prime$ self-adjoint and the 
sequence $(L^\prime_\ell)_{\ell\ge 1}$ is finite or infinite then 
the conditions {\rm (i)--(iv)} are fulfilled with $H,(L_\ell)_{\ell\ge 1}$ 
replaced by $H^\prime,(L^\prime_\ell)_{\ell\ge 1}$ respectively if 
and only if the lengths of the sequences $(L_\ell)_{\ell\ge 1}$, 
$(L^\prime_\ell)_{\ell\ge 1}$ are equal and for some scalar 
$c\in\mathbb{R}$ and a unitary matrix $(u_{\ell j})_{\ell,j}$ we have 
\[
H^\prime=H+c,\qquad L^\prime_\ell=\sum_{j}u_{\ell j}L_j.
\]
\end{theorem}

As an immediate consequence of the uniqueness (up to a scalar) of 
the Hamiltonian $H$, the decomposition of $\Ll$ as the sum of the 
derivation $i[H,\cdot]$ and a dissipative part $\Ll_0=\Ll-i[H,\cdot\,]$ 
determined by special GKSL representations of $\Ll$ is unique. 
Moreover, since $(u_{\ell j})$ is unitary, we have 
\[
\sum_{\ell\ge 1}\left(L^\prime_\ell\right)^*L^\prime_\ell 
=\sum_{\ell,k,j\ge 1}\overline{u}_{\ell k}u_{\ell j}L_k^*L_j 
=\sum_{k,j\ge 1}
\left(\sum_{\ell\ge 1}\overline{u}_{\ell k}u_{\ell j}\right) L_k^*L_j 
=\sum_{k\ge 1}L_k^*L_k. 
\]
Therefore, putting $G=-2^{-1}\sum_{\ell\ge 1}L^*_\ell L_\ell -iH$, 
we can write $\Ll$ in the form (\ref{eq-GKSL-G}) where $G$ is 
uniquely determined by $\Ll$ up to a purely imaginary multiple 
of the identity operator. 

Theorem \ref{th-special-GKSL} can be restated in the 
index free form (\cite{Partha} Thm. 30.12).

\begin{theorem}\label{th-special-GKSL-k} 
Let $\Ll$ be the generator of a uniformly continuous QMS on $\Bh$, 
then there exist an Hilbert space $\mathsf{k}$, a bounded linear 
operator $L:\h\rightarrow\h\otimes \mathsf{k}$ and a bounded 
self-adjoint operator $H$ in $\h$ satisfying the following:
\begin{enumerate}
\item $\Ll(x)=i[H,x]-\frac{1}{2}\left(L^*Lx
        -2L^*(x\otimes{\hbox{\em 1\kern-2.8truept l}}_\mathsf{k})L+xL^*L\right)$ 
         for all $x\in{\mathcal{B}}(\h)$;
\item the set 
      $\{(x\otimes{\hbox{\em 1\kern-2.8truept l}}_\mathsf{k})
      Lu:x\in{\mathcal{B}}(\h),\ u\in\h\}$ 
      is total in $\h\otimes \mathsf{k}$.
\end{enumerate}
\end{theorem}

\noindent{\bf Proof.} 
Let $\mathsf{k}$ be a Hilbert space with Hilbertian dimension 
equal to the length of the sequence $(L_k)_k$ and let $(f_k)$ 
be an orthonormal basis of $\mathsf{k}$. 
Defining $Lu=\sum_kL_ku\otimes f_k$, where $(f_k)$ is 
an orthonormal basis of $\mathsf{k}$ and the $L_k$ are as 
in Theorem \ref{th-special-GKSL}, a simple calculation shows 
that 1 is fulfilled.

Suppose that there exists a non-zero vector $\xi$ orthogonal 
to the set of $(x\otimes\unit_\mathsf{k})Lu$ with 
$x\in{\mathcal{B}}(\h),\ u\in\h$; then 
$\xi=\sum_kv_k\otimes f_k$ with $v_k\in\h$ and 
\[ 
0 = \langle \xi,(x\otimes\unit_\mathsf{k})Lu\rangle 
  = \sum_k\langle v_k,xL_ku\rangle 
  = \sum_k\langle L^*_kx^*v_k,u\rangle 
\] 
for all $x\in{\mathcal{B}}(\h)$, $u\in\h$. Hence, 
$\sum_kL^*_kx^*v_k=0$. Since $\xi\not=0$, we can suppose 
$\Vert v_1\Vert=1$; then, putting $p=|{v_1}\rangle\langle{v_1}|$ 
and $x=py^*$, $y\in{\mathcal{B}}(\h)$, we get 
\begin{equation}\label{eq-noli}
0=L^*_1yv_1 +\sum_{k\geq 2}\langle v_1,v_k\rangle L^*_kyv_1
=\bigg(L^*_1
+\sum_{k\geq 2}\langle v_1,v_k\rangle L^*_k\bigg)yv_1.
\end{equation} 
Since $y\in{\mathcal{B}}(\h)$ is arbitrary, equation 
(\ref{eq-noli}) contradicts the linear independence of 
the $L_k$'s. Therefore the set in (2) must be total. \qed  
\smallskip

The Hilbert space $\mathsf{k}$ is called the 
{\em multiplicity space} of the completely positive part of $\Ll$. 
A unitary matrix $(u_{\ell j})_{\ell,j\ge 1}$, in the above 
basis $(f_k)_{k\ge 1}$, clearly defines a unitary operator on 
$\mathsf{k}$. From now on we shall identify such matrices with 
operators on $\mathsf{k}$.

We end this section by establishing the relationship between  
the operators $G,L_\ell$ and $G^\prime, L_\ell^\prime$ in two  
special GKSL representations of $\Ll$ and $\Ll^\prime$ when 
these generator are both bounded. 

The dual QMS $\T^\prime$ clearly satisfies
\[
\rho^{1/2}\T^\prime_t(x)\rho^{1/2} 
= \T_{*t}(\rho^{1/2} x\rho^{1/2})
\] 
where $\T_*$ denotes the predual semigroup of $\T$. Since  
$\Ll^\prime$ is bounded, differentiating at $t=0$, 
we find the relationship among the generator $\Ll^\prime$ of $\T$ 
and $\Ll_*$ of the predual semigroup $\T_{*}$ of $\T$
\begin{equation}\label{eq-duality-generators} 
\rho^{1/2}\Ll^\prime(x)\rho^{1/2}=\Ll_*(\rho^{1/2} x \rho^{1/2}).
\end{equation}

\begin{proposition}\label{prop-G-from-L} 
Let $\Ll(a)=G^*a+aG+\sum_\ell L^*_\ell aL_\ell$ be a special GKSL 
representation of $\Ll$ with respect to  a $\T$-invariant state 
$\rho=\sum_k \rho_k |e_k\rangle\langle e_k|$. Then 
\begin{eqnarray}
G^*u &=&\sum_{k\geq 1}\rho_k\Ll(\ketbra{u}{e_k})e_k-\hbox{\em tr}(\rho G)u \label{eq-G-star-u}\\
 Gv  &=&\sum_{k\geq 1}\rho_k\Ll_*(\ketbra{v}{e_k})e_k-\hbox{\em tr}(\rho G^*)v\label{eq-G-v}
\end{eqnarray} 
for every $u,v\in\h$.
\end{proposition}

\noindent{\bf Proof.} 
Since $\Ll(|{u}\rangle\langle{v}|) =|{G^*u}\rangle\langle{v}|
+|u\rangle\langle{Gv}|+\sum_\ell |{L^*_\ell u}\rangle\langle{L^*_\ell v}|$, 
putting $v=e_k$ we have $G^*u =|{G^*u}\rangle\langle{e_k}|e_k$ and 
\[
G^*u =\Ll(|u\rangle\langle e_k|)e_k  
- \sum_{\ell}\langle e_k,L_\ell e_k\rangle L^*_\ell u 
- \langle  e_k,Ge_k\rangle u.
\] 
Multiplying both sides by $\rho_k$ and summing on $k$, we find then 
\begin{eqnarray*}
G^*u&=&\sum_{k\geq 1}\rho_k\Ll(|u\rangle\langle e_k|)e_k
 - \sum_{\ell,k}\rho_k\langle e_k,L_\ell e_k\rangle L^*_\ell u
-\sum_{k\geq 1}\rho_k\langle  e_k,Ge_k\rangle u\\
&=&\sum_{k\geq 1}\rho_k\Ll(\ketbra{u}{e_k})e_k
-\sum_\ell\tr(\rho L_\ell)L^*_\ell u-\tr(\rho G)u
\end{eqnarray*} 
and (\ref{eq-G-star-u}) follows since $\tr(\rho L_j)=0$. 
The identity (\ref{eq-G-v}) is now immediate computing the 
adjoint of $G$.  \qed

\begin{proposition}\label{prop-G-and-Gprime} 
Let $\T^\prime$ be the dual of a QMS $\T$ generated by 
${\Ll}$ with normal invariant state $\rho$. 
If $G$ and ${G}^\prime$ are the operators (\ref{eq-G-v}) 
in two GKSL representations of $\Ll$ and 
${\Ll}^\prime$ then 
\begin{equation}\label{eq-G-and-Gprime}
G^\prime\rho^{1/2}=\rho^{1/2}G^*
         +\left(\hbox{\em tr}(\rho G)
         -\hbox{\em tr}(\rho {G}^\prime)\right)\rho^{1/2}.
\end{equation} 
Moreover, we have 
$\hbox{\em tr}(\rho G)-\hbox{\em tr}(\rho{G}^\prime)=ic$ 
for some $c\in\mathbb{R}$.
\end{proposition}

\noindent{\bf Proof.} 
The identities (\ref{eq-G-v}) and (\ref{eq-duality-generators}) yield 
\begin{eqnarray*} 
{G}^\prime\rho^{1/2}
 &=&\sum_{k\geq 1}{\Ll}_*^\prime 
   (\rho^{1/2}\ketbra{v}{\rho_k^{1/2}e_k})\rho_k^{1/2} e_k-
     \tr(\rho {G}^{\prime*})\rho^{1/2}v\\
 &=&\sum_{k\geq 1}
   {\Ll}_*^\prime(\rho^{1/2}(\ketbra{v}{e_k})\rho^{1/2})  
     \rho^{1/2} e_k-\tr(\rho {G}^{\prime*})\rho^{1/2}v\\  
 &=&\sum_{k\geq 1}\rho^{1/2}\Ll(\ketbra{v}{e_k})\rho^{1/2}\rho^{1/2} e_k 
     -\tr(\rho {G}^{\prime*})\rho^{1/2}v\\
 &=&\rho^{1/2}G^*v+\left(\tr(\rho G)
   -\tr(\rho {G}^{\prime*})\right)\rho^{1/2}v.
\end{eqnarray*} Therefore, we obtain (\ref{eq-G-and-Gprime}). 
Right multiplying this equation by $\rho^{1/2}$ we have 
${G}^\prime\rho=\rho^{1/2}G^*\rho^{1/2}+\left(\tr(\rho G) 
-\tr(\rho{G}^{\prime*})\right)\rho$, 
and, taking the trace,
\begin{eqnarray*}
\tr(\rho G)-\tr(\rho{G}^{\prime*})
&=&\tr({G}^\prime\rho)-\tr(\rho^{1/2}G^*\rho^{1/2})\\
&=&\tr({G}^\prime\rho)-\tr(G^*\rho)
=-\overline{\left(\tr(\rho G)-\tr(\rho{G}^{\prime*})\right)};
\end{eqnarray*} 
this proves the last claim. \qed

We can now prove as in \cite{FFVU} Th. 7.2 p. 358 the following

\begin{theorem}\label{th-dualesimm} 
For all special GKSL representation of $\Ll$ by means of operators
$G,L_\ell$ as in (\ref{eq-GKSL}) there exists a special GKSL 
representation of $\Ll^\prime$ by means of operators 
$G^\prime,L^\prime_\ell$ such that:
\begin{enumerate} 
\item\label{cnd-symm-DB-1inf} 
     $G^\prime\rho^{1/2}=\rho^{1/2}G^*+ic\rho^{1/2}$ for some 
     $c\in\mathbb{R}$,
\item\label{cnd-symm-DB-2inf} 
     $L^\prime_\ell\,\rho^{1/2}=\rho^{1/2}L^*_\ell$.
\end{enumerate}
\end{theorem}

\noindent{\bf Proof.} 
Since $\Ll^\prime$ is bounded, it admits a special GKSL representation 
$\mathcal{L}^\prime(a)=G^{\prime *}a+\sum_kL^{\prime *}_kaL^\prime_k+aG^\prime$.
Moreover, by Proposition \ref{prop-G-and-Gprime} we have
$G^\prime\rho^{1/2}=\rho^{1/2}G^*+ic$, $c\in\mathbb{R}$, and so 
(\ref{eq-duality-generators}) implies
\begin{equation}\label{partecp}
\sum_k\rho^{1/2}L^{\prime *}_k x L^\prime_k\rho^{1/2}=
\sum_kL_k\rho^{1/2}x\rho^{1/2}L^*_k.
\end{equation}
Let $\mathsf{k}$ (resp. $\mathsf{k}'$) be the multiplicity space 
of the completely positive part of $\Ll$ (resp. $\Ll'$) and define 
an operator $X:\h\otimes\mathsf{k}'\to\h\otimes\mathsf{k}$ 
\[
X(x\otimes\unit_{\mathsf{k}^\prime})L^\prime\rho^{1/2}u
=(x\otimes\unit_{\mathsf{k}})(\rho^{1/2}\otimes\unit_\mathsf{k})L^*u
\] 
for all $x\in\Bh$ and $u\in\h$, where 
$L:\h\rightarrow\h\otimes \mathsf{k}$, 
$Lu=\sum_kL_ku\otimes f_k$, 
$L^\prime:\h\rightarrow\h\otimes \mathsf{k}^\prime$, 
$L^\prime u=\sum_kL^\prime_ku\otimes f^\prime_k$, 
$(f_k)_k$ and $(f^\prime_k)_k$ are orthonormal bases 
of $\mathsf{k}$ and $\mathsf{k}^\prime$ respectively.
Thus, by (\ref{partecp}),
\begin{eqnarray*}
& & \langle X(x\otimes\unit_{\mathsf{k}^\prime})L^\prime\rho^{1/2} u, 
X(y\otimes\unit_{\mathsf{k}^\prime})L^\prime \rho^{1/2}v\rangle
=\sum_k\langle u,\rho^{1/2}L^{\prime *}_k x^*yL^{\prime}_k \rho^{1/2}v\rangle \\
& & \hskip 3truecm =\langle(x\otimes\unit_{\mathsf{k}^\prime})
L^\prime \rho^{1/2}u,(y\otimes\unit_{\mathsf{k}^\prime})L^\prime \rho^{1/2}v\rangle
\end{eqnarray*}
for all $x,y\in\Bh$ and $u,v\in\h$, i.e. $X$ preserves the 
scalar product.
Therefore, since the set $\{(x\otimes\unit_{\mathsf{k}^\prime})
L^\prime\rho^{1/2} u\,\mid\,x\in\Bh,\ u\in\h\}$ is total 
in $\h\otimes\mathsf{k}^\prime$ (for $\rho^{1/2}(\h)$ is 
dense in $\h$ and Theorem \ref{th-special-GKSL-k} holds), 
we can extend $X$ to an unitary operator from 
$\h\otimes \mathsf{k}^\prime$ to $\h\otimes\mathsf{k}$.
As a consequence we have $X^*X=\unit_{\h\otimes \mathsf{k}^\prime}$.

Moreover, since
$X(y\otimes\unit_{\mathsf{k}^\prime})=(y\otimes\unit_{\mathsf{k}^\prime})X$
for all $y\in\Bh$, we can conclude that $X=\unit_\h\otimes Y$ for
some unitary map
$Y:\mathsf{k}^\prime\rightarrow\mathsf{k}^\prime$. 

The definition of $X$ implies then 
\[
(\rho^{1/2}\otimes\unit_{\mathsf{k}})L^*=XL^\prime\rho^{1/2}
=(\unit_\h\otimes Y)L^\prime\rho^{1/2}.
\]

This means that, replacing $L^\prime$ by $(\unit_\h\otimes Y)L^\prime$, 
or more precisely $L^\prime_k$ by $\sum_{\ell}u_{k\ell}L^\prime_\ell$ for 
all $k$, we have 
\[
\rho^{1/2}L^*_k=L^\prime_k\rho^{1/2}. 
\]

Since $\tr(\rho {L^\prime}_k)= \tr(\rho L_k^*)=0$ and, 
from ${\Ll}^\prime(\unit)=0$, ${G^\prime}^*+ 
{G^\prime}=-\sum_k {L^\prime}^*_k {L^\prime}_k$,  
the properties of a special GKSL representation follow. 
\qed

\begin{remark}\label{rem-cp-parts}
{\rm Condition \ref{cnd-symm-DB-2inf} implies that the 
completely positive parts $\varPhi(x)=\sum_\ell L^*_\ell x L_\ell$ 
and $\varPhi^\prime $ of the generators $\Ll$ and $\Ll^\prime$, 
respectively are mutually adjoint i.e. 
\begin{equation}\label{Phi-simm}
\tr(\rho^{1/2}\Phi^\prime(x)\rho^{1/2}y)=\tr(\rho^{1/2}x\rho^{1/2}\Phi(y))
\end{equation}
for all $x,y\in\Bh$. As a consequence, also the maps 
$x\to G^*x + x G$ and $x\to (G^\prime)^*x + x G^\prime $ are mutually adjoint.}
\end{remark}

\section{Generators of Standard Detailed Balance QMSs}\label{sect-SDB-QMS} 

In this section we characterise the generators of norm-continuous 
QMSs satisfying the SQDB of Definition \ref{def-SQDB}. 

We start noting that, since $\rho$ is invariant for $\T$ and 
$\T^\prime$, i.e. $\Ll_*(\rho)=\Ll_*^\prime(\rho)=0$, the 
operator $K$ commutes with $\rho$. Moreover, by comparing 
two special GKSL representations of $\Ll$ and 
$\Ll^\prime + 2i[K, \cdot\,]$, we have immediately the following 
 
\begin{lemma}\label{lem-confronto}
A QMS $\mathcal{T}$ satisfies the SQDB $\Ll-\pL=2i[K,\cdot\,]$ 
if and only if for all special GKSL representations of the generators 
$\Ll$ and $\Ll^\prime$ by means of operators $G,L_k$ and 
$G^\prime,L_k^\prime$ respectively, we have
\[
G=G^\prime+2iK+ic\quad\quad
L^\prime_k = \sum_j u_{kj}L_j
\]
for some $c\in\mathbb{R}$ and some unitary 
$(u_{kj})_{kj}$ on $\mathsf{k}$.
\end{lemma}

Since we know the relationship between the operators 
$G^\prime, L_k^\prime$ and $G,L_k$ thanks to Theorem 
\ref{th-dualesimm}, we can now characterise generators 
of QMSs satisfying the SQDB. We emphasize the following 
definition of {\em $T$-symmetric} matrix (operator) on $\mathsf{k}$ 
in order to avoid confusion with the usual notion of symmetric 
operator $X$ meaning that $X^*$ is an extension of $X$. 

\begin{definition}\label{def-T-symm}
Let $Y=(y_{k\ell})_{k,\ell\ge 1}$ be a matrix with entries 
indexed by $k,\ell$ running on the set (finite or infinite) 
of indices of the sequence $(L_\ell)_{\ell\ge 1}$.  We denote 
by $Y^T$ the transpose matrix $Y^T=(y_{\ell k})_{k,\ell\ge 1}$. 
The matrix $Y$ is called {\em $T$-symmetric} if $Y=Y^T$.
\end{definition}

\begin{theorem}\label{th-SQDB}
$\T$ satisfies the SQDB if and only if for all special 
GKSL representation of the generator $\Ll$ by means of 
operators $G,L_k$ there exists a $T$-symmetric unitary 
$(u_{m\ell})_{m\ell}$ on $\mathsf{k}$ such that 
\begin{equation}\label{cond-2}
\rho^{1/2}L^*_k=\sum_\ell u_{k\ell}L_\ell\rho^{1/2},
\end{equation}
for all $k\ge 1$.
\end{theorem}

\noindent{\bf Proof.} Given a special GKSL of $\Ll$, 
Theorem \ref{th-dualesimm} allows us to write the dual 
$\Ll^\prime$ in a special GKSL representation by means of 
operators $G^\prime$, $L^\prime_k$ with  
\begin{equation}\label{eq-Gprime-Lprime}
G^\prime\rho^{1/2}=\rho^{1/2}G^*,\quad\quad L^\prime_k\rho^{1/2}
=\rho^{1/2}L^*_k.
\end{equation}

Suppose first that $\T$ satisfies the SQDB. Since 
$L^\prime_k = \sum_j u_{kj}L_j$ for some unitary  
$(u_{kj})_{kj}$ by Lemma \ref{lem-confronto}, we can find 
(\ref{cond-2}) substituting $L^\prime_k$ with $\sum_j u_{kj}L_j$
in the second formula (\ref{eq-Gprime-Lprime}). 

Finally we show that the unitary matrix $u=(u_{m\ell})_{m\ell}$ 
is $T$-symmetric. Indeed, taking the adjoint of (\ref{cond-2}) we 
find $L_\ell\rho^{1/2}=\sum_m \bar{u}_{\ell m}\rho^{1/2}L_m^*$. 
Writing $\rho^{1/2}L_m^*$ as in (\ref{cond-2}) we have then
\[
L_\ell\rho^{1/2}=\sum_{m,k} \bar{u}_{\ell m} u_{mk} L_k \rho^{1/2}
= \sum_k \left((u^*)^T u\right)_{\ell k}L_k \rho^{1/2}. 
\] 
The operators $L_\ell\rho^{1/2}$ are linearly independent by property (iii) 
Theorem \ref{th-special-GKSL} of a special GKSL representation, 
therefore $(u^*)^T u$ is the identity operator on $\mathsf{k}$. 
Since $u$ is also unitary, we have also $u^*u=(u^*)^T u$, namely  
$u^*= (u^*)^T$ and $u=u^T$.

Conversely, if (\ref{cond-2}) holds, by (\ref{eq-Gprime-Lprime}), 
we have $L^\prime_k\rho^{1/2}=\sum_\ell u_{k\ell}L_\ell\rho^{1/2}$, 
so that $L^\prime_k=\sum_\ell u_{k\ell}L_\ell$ for all $k$ and for 
some unitary  $(u_{kj})_{kj}$.
Therefore, thanks to Lemma \ref{lem-confronto}, to conclude it is 
enough to prove that $G=G^\prime+i(2K+c)$ namely, that $G-G^\prime$ 
is anti self-adjoint.

To this end note that, since $\rho$ is an invariant state, we have
\begin{equation}\label{rhoinvG}
0=\rho G^*+\sum_kL_k\rho L^*_k+G\rho,
\end{equation}
with
\begin{eqnarray*}
\sum_kL_k\rho L^*_k &=& \sum_k(L_k\rho^{1/2})(\rho^{1/2} L^*_k)
=\sum_k\sum_{\ell,j} 
    \overline{u}_{k\ell}u_{kj}\rho^{1/2}L^*_\ell L_j\rho^{1/2}\\
&=&\sum_\ell\rho^{1/2}L_\ell^*L_\ell\rho^{1/2}
=-\rho^{1/2}(G+G^*)\rho^{1/2},
\end{eqnarray*} 
(for condition (\ref{cond-2}) holds) and so, by substituting in 
equation $(\ref{rhoinvG})$ we get
\begin{eqnarray*}
0 &=&\rho G^*-\rho^{1/2}G\rho^{1/2}-\rho^{1/2}G^*\rho^{1/2}+G\rho
   =\rho^{1/2}\left(\rho^{1/2}G^*-G\rho^{1/2}\right) \\
  &-&\left(\rho^{1/2}G^*-G\rho^{1/2}\right)\rho^{1/2}
   =[G\rho^{1/2}-\rho^{1/2}G^*,\rho^{1/2}],
\end{eqnarray*}
i.e. $G\rho^{1/2}-\rho^{1/2}G^*$ commutes with $\rho^{1/2}$. 

We can now prove that $G-G^\prime$ is anti self-adjoint. Clearly, 
it suffices to show that 
$\rho^{1/2}G\rho^{1/2}-\rho^{1/2}G^\prime\rho^{1/2}$ is 
anti self-adjoint. Indeed, by (\ref{eq-Gprime-Lprime}), we have 
\begin{eqnarray*}
\left(\rho^{1/2}G\rho^{1/2}-\rho^{1/2}G^\prime\rho^{1/2}\right)^*
& = & \left(\rho^{1/2}G\rho^{1/2}-\rho G^*\right)^* \\
& = & \left(\rho^{1/2}\left(G\rho^{1/2}-\rho^{1/2} G^*\right)\right)^* \\
& = & \left(\left(G\rho^{1/2}-\rho^{1/2} G^*\right)\rho^{1/2}\right)^* \\
& = & \rho G^* - \rho^{1/2}G \rho^{1/2} 
  = \rho^{1/2} G^\prime\rho^{1/2} - \rho^{1/2}G \rho^{1/2}
\end{eqnarray*} 
because $G\rho^{1/2}-\rho^{1/2}G^*$ commutes with $\rho^{1/2}$. 
This completes the proof. \qed 

\smallskip
It is worth noticing that, as in Remark \ref{rem-cp-parts}, $\T$ 
satisfies the SQDB if and only if the completely positive 
part $\varPhi$ of the generator $\Ll$ is symmetric. This 
improves our previous result, Thm. 7.3 \cite{FFVU}, where we 
gave $G\rho^{1/2}=\rho^{1/2}G^* -\left(2i K +ic\right)\rho^{1/2}$ 
for some $c\in\mathbb{R}$ as an additional condition. Here we showed 
that it follows from (\ref{cond-2}) and the invariance of $\rho$. 

\begin{remark}{\rm 
Note that (\ref{cond-2}) holds for the operators $L_\ell$ of a 
special GKSL representation of $\Ll$ if and only if it is true 
for {\em all} special GKSL representations because of the second 
part of Theorem \ref{th-special-GKSL}. Therefore the conclusion 
of Theorem \ref{th-SQDB} holds true also if and only if we can 
find a single special GKSL representation of $\Ll$ satisfying 
(\ref{cond-2}).}
\end{remark}

\smallskip
The $T$-symmetric unitary $(u_{m\ell})_{m\ell}$ is determined 
by the $L_\ell$'s because they are linearly independent. We shall now 
exploit this fact to give a more geometrical characterisation of SQDB. 

When the SQDB holds, the matrices $(b_{kj})_{k,j\ge 1}$ 
and $(c_{kj})_{k,j\ge 1}$ with 
\begin{equation}\label{eq-ops-B&C-on-k}
b_{k j}=\tr\left(\rho^{1/2}L^*_k\rho^{1/2}L_j^*\right), 
\quad\hbox{\rm and }\quad
c_{k j}=\tr\left(\rho L^*_k L_j\right)
\end{equation}
define two trace class operators $B$ and $C$ on  $\mathsf{k}$ 
by Lemma \ref{lem-2bases} (see the Appendix); $B$ is 
$T$-symmetric and $C$ is self-adjoint. Moreover, it admits 
a self-adjoint inverse $C^{-1}$ because $\rho$ is faithful. When 
$\mathsf{k}$ is infinite dimensional, $C^{-1}$ is unbounded and 
its domain coincides with the range of $C$.

We can now give the following characterisation of QMS satisfying the SQDB 
condition which is more direct because the unitary $(u_{k\ell})_{k\ell}$ 
in Theorem \ref{th-SQDB} is explicitly given by $C^{-1}B$. 
 
\begin{theorem}\label{th-SQDB-geo}
$\T$ satisfies the SQDB if and only if the operators $G,L_k$ of 
a special GKSL representation of the generator $\Ll$ satisfy the 
following conditions:
\begin{enumerate}
\item[(i)] the closed linear span of 
$\left\{\rho^{1/2}L_\ell^*\,\mid\,\ell\ge 1\right\}$ 
and $\left\{L_\ell \rho^{1/2}\,\mid\,\ell\ge 1\right\}$ in the 
Hilbert space of Hilbert-Schmidt operators on $\mathsf{h}$ coincide,
\item[(ii)] the trace-class operators $B,C$ defined by 
(\ref{eq-ops-B&C-on-k}) satisfy 
$C B = B C^T$ and $C^{-1}B$ is unitary $T$-symmetric. 
\end{enumerate}
\end{theorem}

\noindent{\bf Proof.} If $\T$ satisfies the SQDB then, by 
Theorem \ref{th-SQDB}, the identity (\ref{cond-2}) holds. 
The series in the right-hand side of (\ref{cond-2}) is 
convergent with respect to the Hilbert-Schmidt norm because 
\begin{eqnarray*}
\left\Vert \sum_{m+1\le \ell\le n} u_{k\ell}L_\ell\rho^{1/2}\right\Vert_{HS}^2 
& = & \sum_{m+1\le\ell,\ell^\prime\le n} 
\bar{u}_{k\ell^\prime}u_{k\ell}\tr\left(\rho L^*_{\ell^\prime} L_\ell\right) \\
& \le & 
\frac{1}{2}\sum_{m+1\le\ell,\ell^\prime\le n} 
|u_{k\ell^\prime}|^2\,|u_{k\ell}|^2 
+\frac{1}{2}\sum_{m+1\le\ell,\ell^\prime\le n} 
|c_{\ell^\prime \ell}|^2 \\
& \le & 
\frac{1}{2}\left(\sum_{m+1\le\ell\le n} |u_{k\ell}|^2\right)^2 
+\frac{1}{2}\sum_{m+1\le\ell,\ell^\prime\le n} 
|c_{\ell^\prime \ell}|^2 
\end{eqnarray*}
and the right-hand side vanishes as $n,m$ go to infinity because 
the operator $C$ is trace-class by Lemma \ref{lem-2bases} and 
the columns of $U=(u_{k\ell})_{k\ell}$ are unit vectors in $\mathsf{k}$ 
by unitarity.

Left multiplying both sides of (\ref{cond-2}) by $\rho^{1/2}L^*_j$ 
and taking the trace we find $B=CU^T=CU$. It follows that the range 
of the operators $B$, $CU$ and $C$ coincide and $C^{-1}B=U$ is 
everywhere defined, unitary and $T$-symmetric because $U$ is 
$T$-symmetric. Moreover, since $B$ is $T$-symmetric by the cyclic 
property of the trace, we have also 
\[
BC^T = CU^TC^T = C(CU)^T=CB^T=CB.
\]

Conversely, we show that (i) and (ii) imply the SQDB. To this end 
notice that, by the spectral theorem we can find a unitary linear 
transformation $V=(v_{mn})_{m,n\ge 1}$ on $\mathsf{k}$ such that $V^*CV$ 
is diagonal. Therefore, choosing a new GKSL of the generator $\Ll$ by 
means of the operators ${L''}_k = \sum_{n\ge 1}v_{nk}L_n$, if necessary, 
we can suppose that both $(L_\ell\rho^{1/2})_{\ell\ge 1}$ and 
$(\rho^{1/2} L_k^*)_{k\ge 1}$ are {\em orthogonal} bases of the 
same closed linear space. Note that 
\[
\tr(\rho^{1/2}(L'')^*_k\rho^{1/2}(L'')^*_j) 
=\sum_{m,n\ge 1}\bar{v}_{nk}\bar{v}_{mj} 
\tr(\rho^{1/2}L^*_n\rho^{1/2}L^*_m)
\]
and the operator $B$, after this change of GKSL representation, 
becomes $V^*B (V^*)^T$ which is also $T$-symmetric.

Writing the expansion of $\rho^{1/2}L_k^*$ with respect to the 
orthogonal basis $(L_\ell\rho^{1/2})_{\ell\ge 1}$, for all $k\ge 1$ 
we have 
\begin{equation}\label{eq-expansionLk}
\rho^{1/2}L_k^* = \sum_{\ell\ge 1}
\frac{\tr(\rho^{1/2}L^*_\ell\rho^{1/2}L^*_k)}
{\Vert L_\ell\rho^{1/2}\Vert_{HS}^2}\,L_\ell\,\rho^{1/2}.
\end{equation}
In this way we find a matrix $Y$ of complex numbers $y_{k\ell}$ 
such that $\rho^{1/2}L_k^*=\sum_{\ell} y_{k\ell} L_\ell\rho^{1/2}$ 
and the series is Hilbert-Schmidt norm convergent. Clearly, since 
$C$ is diagonal and $B$ is $T$-symmetric, 
$y_{k\ell}=(BC^{-1})_{k\ell}=((B(C^{-1})^T)_{k\ell}=((C^{-1}B)^T)_{k\ell}$.
It follows from (ii) that $Y$ coincides with the unitary operator 
$(C^{-1}B)^T$ and (\ref{cond-2}) holds. Moreover, $Y$ is symmetric 
because  
\[
y_{\ell k}=(BC^{-1})_{\ell k}  
= ((B(C^{-1})^T)_{\ell k} = (C^{-1}B)_{k\ell} = y_{k\ell}.
\]
This completes the proof. \qed 

\medskip
Formula (\ref{eq-expansionLk}) has the following consequence.

\begin{corollary}\label{cor-weightLl=weightLk}
Suppose that a QMS $\T$ satisfies the SQDB condition. 
For every special GKSL representation of $\Ll$ with operators 
$L_\ell\rho^{1/2}$ that are orthogonal in the Hilbert space of 
Hilbert-Schmidt operators on $\mathsf{h}$ if 
$\tr(\rho^{1/2}L^*_\ell\rho^{1/2}L^*_k)\not=0$  
for a pair of indices $k,\ell\ge 1$, then
$\tr(\rho L_\ell^*L_\ell)=\tr(\rho L_k^*L_k)$.
\end{corollary}

\noindent{\bf Proof.} It suffices to note that the matrix 
$(u_{k\ell})$ with entries 
\[
u_{k\ell}=\frac{\tr(\rho^{1/2}L^*_\ell\rho^{1/2}L^*_k)}
{\Vert L_\ell\rho^{1/2}\Vert_{HS}^2}
=\frac{\tr(\rho^{1/2}L^*_\ell\rho^{1/2}L^*_k)}
{\tr(\rho L^*_\ell L_\ell)}
\] 
must be $T$-symmetric. \qed

\begin{remark}\label{rem-covariance}
{\rm
The matrix $C$ can be viewed as the covariance matrix of the zero-mean
(recall that $\tr(\rho L_\ell)=0$) ``random variables'' 
$\{\,L_\ell\,\mid\,\ell\ge 1\,\}$ and in a similar way, $B$ can be 
viewed as a sort of mixed covariance matrix between the previous random 
variable and the adjoint $\{\,L_\ell^*\,\mid\,\ell\ge 1\,\}$. 
Thus the SQDB condition holds when the random variables $L_\ell$ right 
multiplied by $\rho^{1/2}$ and the adjoint variables $L_\ell^*$ 
left multiplied by $\rho^{1/2}$ generate the same subspace of 
Hilbert-Schmidt operators and the mixed covariance matrix $B$ is 
a left unitary transformation of the covariance matrix $C$.

If we consider a special GKSL representation of $\Ll$ with operators 
$L_\ell\rho^{1/2}$ that are orthogonal, then, by 
Corollary \ref{cor-weightLl=weightLk} and the identity 
$\Vert L_\ell\rho^{1/2}\Vert_{HS}= \Vert L_k\rho^{1/2}\Vert_{HS}$, 
the unitary matrix $U$ can be written as $C^{-1/2}B C^{-1/2}$. This, 
although not positive definite, can be interpreted as a {\
em correlation coefficient} matrix of $\{\,L_\ell\,\mid\,\ell\ge 1\,\}$ 
and $\{\,L_\ell^*\,\mid\,\ell\ge 1\,\}$.} 
\end{remark} 

\medskip

The characterisation of generators of symmetric QMSs with respect to the 
$s=1/2$ scalar product follows along the same lines.

\begin{theorem}\label{th-KMSsimm} 
A norm-continuous QMS $\T$ is symmetric if and only if 
there exists a special GKSL representation of the generator 
$\Ll$ by means of operators $G,L_\ell$ such that 
\begin{itemize}
\item[\rm(1)] 
 $G\rho^{1/2}=\rho^{1/2}G^* +ic\rho^{1/2}$ for some $c\in\mathbb{R}$,
\item[\rm(2)]
     $\rho^{1/2}L^*_k=\sum_\ell u_{k\ell}L_\ell\rho^{1/2}$, 
for all $k$, for some unitary  $(u_{k\ell})_{k\ell}$ on $\mathsf{k}$
which is also $T$-symmetric.
\end{itemize}
\end{theorem}

\noindent{\bf Proof.}  Choose a special GKSL representation of 
$\Ll$ by means of operators $G$, $L_k$. Theorem \ref{th-dualesimm} 
allows us to write the symmetric dual $\Ll^\prime$ in a special GKSL 
representation by means of operators $G^\prime$, $L^\prime_k$ 
as in (\ref{eq-Gprime-Lprime}).

Suppose first that $\T$ is KMS-symmetric. Comparing the 
special GKSL representations of $\Ll$ and $\Ll^\prime$, by 
Theorem \ref{th-special-GKSL} we find
\[
G = G^\prime+ic, \quad  
L^\prime_k = \sum_j u_{kj}L_j,
\]
for some unitary matrix $(u_{kj})$ and some $c\in\mathbb{R}$. 
This, together with (\ref{eq-Gprime-Lprime}) implies that 
conditions (1) and (2) hold. 

Assume now that conditions (1) and (2) hold. Taking the adjoint 
of (2) we find immediately $L_k\rho^{1/2} 
=\sum_{k}\overline{u}_{k\ell}\rho^{1/2}L^*_\ell$. Then straightforward 
computation, by the unitarity of the matrix $(u_{k\ell})$, yields 
\begin{eqnarray*}
\Ll_*(\rho^{1/2}x\rho^{1/2}) 
& = & G\rho^{1/2}x\rho^{1/2}
       + \sum_k L_k \rho^{1/2} x\rho^{1/2} L^*_k 
       + \rho^{1/2}x\rho^{1/2} G^*\\
& = & \rho^{1/2}G^*x\rho^{1/2}
       + \sum_{\ell\, k j} \overline{u}_{k\ell}\,u_{kj}\, 
              \rho^{1/2}L^*_k x L_j \rho^{1/2}
       + \rho^{1/2}xG\rho^{1/2} \\
& = &  \rho^{1/2}\Ll(x)\rho^{1/2}
\end{eqnarray*}
for all $x\in\Bh$. Iterating we find $\Ll^n_*(\rho^{1/2}x\rho^{1/2}) 
=\rho^{1/2}\Ll^n(x)\rho^{1/2}$ for all $n\ge 0$, therefore, exponentiating,  
we find $\T_{*t}(\rho^{1/2}x\rho^{1/2}) = \rho^{1/2}\T_t(x)\rho^{1/2}$
for all $t\ge 0$. This, together with (\ref{eq-duality-T}), 
implies that $\T$ is KMS-symmetric. \quad\qed 

\begin{remark}\label{rem-Phi-KMS-symm} {\rm
Note that condition (2) in Theorem \ref{th-KMSsimm} implies that the 
completely positive part of $\Ll$ is KMS-symmetric. This makes a 
parallel with Theorem \ref{th-dualesimm}  where condition 
(\ref{cnd-symm-DB-2inf}) implies that the completely 
positive parts of the generators $\Ll$ and $\Ll^\prime$ 
are mutually adjoint.}
\end{remark} 

\medskip 
The above theorem simplifies a previous result by Park (\cite{Park} Thm 2.2) 
where conditions (1) and (2) appear in a much more complicated way.

\section{Generators of Standard Detailed Balance (with time 
reversal) QMSs}\label{sect-SDB-TR-QMS} 

We shall now study generators of semigroups satisfying the 
SQDB-$\theta$ introduced in Definition \ref{def-SQDB-theta} 
involving the time reversal operation. In this section, we always 
assume that the invariant state $\rho$ and the anti-unitary time 
reversal $\theta$ commute. 

The relationship between the QMS satisfying the SQDB-$\theta$, 
its dual and their generators is clarified by the following   

\begin{proposition}\label{prop-TR-SDBsstL} 
A QMS $\mathcal{T}$ satisfies the {SQDB-$\theta$} if and only 
if the dual semigroup $\T^\prime$ is given by
\begin{equation}\label{eq-T-and-Tprime-theta}
\T^\prime_t(x)=\theta{\mathcal{T}_t(\theta x \theta )}\theta
\quad\quad\mbox{for all $x\in\Bh$}.
\end{equation}
In particular, if $\T$ is norm-continuous, then $\T^\prime$ is also 
norm-continuous. Moreover, in this case $\T^\prime$ is generated by 
\begin{equation}\label{eq-L-and-Lprime-theta}
\Ll^\prime(x)=\theta{\mathcal{L}(\theta x \theta )}\theta,
\quad\quad x\in\Bh.
\end{equation}
\end{proposition}

\noindent{\bf Proof.} Suppose that $\T$ satisfies the 
SQDB-$\theta$ and put 
$\sigma(x)=\theta x\theta$. Taking $t=0$ equation 
(\ref{eq-SQDB-theta-T}) reduces to 
$\tr(\rho^{1/2} x\rho^{1/2} y) 
=\tr(\rho^{1/2}\,\sigma(y^*)\rho^{1/2}\sigma(x^*))$ 
for all $x,y\in\Bh$, so that
\begin{eqnarray*}
\tr(\rho^{1/2} x\rho^{1/2}\mathcal{T}_t(y))
&=&\tr(\rho^{1/2}\, \sigma(y^*)\rho^{1/2}\mathcal{T}_t(\sigma(x^*)))\\
&=&\tr(\rho^{1/2}\,\sigma(\mathcal{T}_t(\sigma(x^*))^*\rho^{1/2}
            \sigma(\sigma(y^*)^*))\\
&=&\tr(\rho^{1/2}\,\sigma(\mathcal{T}_t(\sigma(x)))\rho^{1/2}y)
\end{eqnarray*} 
for every $x,y\in\Bh$ and (\ref{eq-T-and-Tprime-theta}) follows. 
Therefore, if $\T$ is norm continuous, 
$\T_t^\prime= (\sigma\circ\T_t\circ\sigma)_t$ is also.

Conversely, if (\ref{eq-T-and-Tprime-theta}) holds, the commutation 
between $\rho$ and $\theta$ implies
\begin{eqnarray*}
\tr(\rho^{1/2} \pT_t(x)\rho^{1/2}y)
&=&\tr\left(\rho^{1/2} \theta\T_t(\theta x\theta)\theta \rho^{1/2}y\right)\\
& = & \tr\left(\theta \left(\rho^{1/2} \T_t(\theta x\theta )
     \theta \rho^{1/2}y\theta\right) \theta \right) \\
& = & \tr\left(\rho^{1/2} \theta y^*\rho^{1/2}\theta  \T_t(\theta x^*\theta )\right)
\end{eqnarray*} 
and (\ref{eq-T-and-Tprime-theta}) is proved. 
Now (\ref{eq-L-and-Lprime-theta}) follows from
(\ref{eq-T-and-Tprime-theta}) differentiating at $t=0$. \qed 

\medskip

We can now describe the relationship between special 
GKSL representations of $\Ll$ and $\Ll^\prime$.

\begin{proposition}\label{prop-HLtildeTR}
If $\mathcal{T}$ satisfies the {SQDB-$\theta$} then, for every 
special GKSL representation of $\Ll$ by means of operators $H, L_k$, 
the operators $H^\prime=-\theta{H}\theta$ and $L^\prime_k= 
\theta{L_k}\theta$ yield a special GKSL representation of $\Ll^\prime$. 
\end{proposition}

\noindent{\bf Proof.} Consider a special GKSL representation of 
$\Ll$ by means of operators $H$, $L_k$. Since $\Ll^\prime (a)= 
\theta{\Ll(\theta{a}\theta)}\theta$ by Proposition \ref{prop-TR-SDBsstL}, from the 
antilinearity of $\theta$ and $\theta^2=\unit$ we get
\begin{eqnarray*}
\theta{\Ll^\prime (a)\,}\theta & = &i[H,\theta{a}\theta]
-\frac{1}{2}\sum_k\left(L^*_kL_k\theta{a}\theta-2L^*_k\theta{a}\theta L_k 
         +\theta{a}\theta L^*_kL_k\right)\\
&=&i\trhi{\left(\trhi{H}a-a\trhi{H}\right)}
         +\sum_k\trhi{\left((\trhi{L^*_k})a(\trhi{L_k})\right)}\\
&-&\frac{1}{2}\sum_k\trhi{\left((\trhi{L^*_k})(\trhi{L_k})a
         +a(\trhi{L^*_k})(\trhi{L_k})\right)}\\
&=&\trhi{\left(-i[\trhi{H},a]\,\right)} 
         -\frac{1}{2}\sum_k\trhi{\left(L^{\prime*}_kL^\prime_ka
                  -2L^{\prime *}_kaL^\prime_k
                  +aL^{\prime*}_kL^\prime_k\right)},
\end{eqnarray*}
where $L^\prime_k:=\trhi{L_k}$.
Therefore, putting $H^\prime=-\theta H\theta$, we find a 
GKSL representation of $\Ll^\prime$ which is also special 
because $\tr(\rho L^\prime_k)=\tr(\theta\rho L_k\theta)
=\tr(L_k^*\rho)=\overline{\tr(\rho L_k)}=0$. \qed 

\medskip

The structure of generators of QMSs satisfying 
the {SQDB-$\theta$} is described by the following

\begin{theorem}\label{th-SQDB-TR}
A QMS $\T$ satisfies the {SQDB-$\theta$} condition if and only if 
there exists a special GKSL representation of $\Ll$, with  
operators $G, L_\ell$, such that:
\begin{enumerate}
\item \label{tr-sdb-1} $\rho^{1/2}\theta{G^*}\theta
                =G\rho^{1/2} $, 
\item \label{tr-sdb-2} $\rho^{1/2}\theta{L}_k^*\theta 
      = \sum_j u_{k j} {L_j}\rho^{1/2}$ 
      for a self-adjoint unitary $(u_{k j})_{kj}$ on $\mathsf{k}$.
\end{enumerate}
\end{theorem}
\noindent{\bf Proof.} Suppose that $\T$ satisfies the SQDB-$\theta$ 
condition and consider a special GKSL representation of the generator 
$\Ll$ with operators $G,L_k$. The operators $-\theta{H}\theta$ and 
$\theta{L_k}\theta$ give then a special GKSL representation of 
$\Ll^\prime$ by Proposition \ref{prop-HLtildeTR}. 
Moreover, by Theorem \ref{th-dualesimm}, we have another special GKSL 
representation of $\Ll^\prime$ by means of operators $G^\prime, L^\prime_k$ 
such that $G^\prime\rho^{1/2}=\rho^{1/2}G^*+ic\rho^{1/2}$ for some 
$c\in\mathbb{R},$ and $L^\prime_k\rho^{1/2}=\rho^{1/2}L^*_k$. Therefore 
there exists a unitary $(v_{k j})_{kj}$ on $\mathsf{k}$ such that 
$L^\prime_k=\sum_j v_{kj}\theta L_j\theta$, and 
$\rho^{1/2}L^*_k = \sum_j v_{kj}\theta L_j\theta\rho^{1/2}$.
Condition \ref{tr-sdb-2} follows then with $u_{kj}=\bar{v}_{kj}$ 
left and right multiplying by the antiunitary $\theta$.

In order to find condition \ref{tr-sdb-1}, first notice that 
by the unitarity of $(v_{k j})_{kj}$ 
\begin{equation}\label{Lprimo}
\sum_kL^{\prime *}_kL^\prime_k=\sum_k\theta{L^*_kL_k}\theta.
\end{equation}
Now, by the uniqueness of $G^\prime$ up to a purely imaginary 
multiple of the identity in a special GKSL representation, 
$H^\prime=(G^{\prime *}-G^\prime)/(2i)$ is equal to 
$-\theta H\theta +c_1$ for some $c_1\in\mathbb{R}$. From 
(\ref{Lprimo}) and $G^\prime\rho^{1/2}=\rho^{1/2}G^*+ic\rho^{1/2}$ 
we obtain then 
\begin{eqnarray*}
\rho^{1/2}G^*+ic\rho^{1/2}&=& G^\prime\rho^{1/2} 
  =-iH^\prime\rho^{1/2}
  -\frac{1}{2}\sum_kL^{\prime *}_kL^\prime_k\rho^{1/2} \\ 
 &=&i\theta{H}\theta\rho^{1/2}+i c_1\rho^{1/2}
  -\frac{1}{2}\sum_k\theta{L^*_kL_k}\theta\rho^{1/2} \\
 &=&\theta{G}\theta\rho^{1/2}+i c_1\rho^{1/2}. 
\end{eqnarray*}
It follows that $\rho^{1/2}\theta G^*\theta=G\rho^{1/2}+ic_2\rho^{1/2}$
for some $c_2\in{\mathbb{R}}$. Left multiplying by $\rho^{1/2}$ and tracing 
we find 
\[
c_2 = \tr\left(\theta \rho G^*\theta\right)-\tr(\rho G)=
\tr(G\rho) - \tr(\rho G)=0
\]
and condition $\ref{tr-sdb-1}$ holds. 

Finally we show that the square of the unitary $(u_{kj})_{kj}$ 
on $\mathsf{k}$ is the identity operator. Indeed, taking 
the adjoint of the identity $\rho^{1/2}\theta{L}_k^*\theta 
= \sum_j u_{k j} {L_j}\rho^{1/2}$, we have
\[
\theta L_k\theta\rho^{1/2} = \sum_j \bar{u}_{kj}\rho^{1/2}L_j^*.
\]
Left and right multiplying by the antilinear time reversal $\theta$ 
(commuting with $\rho$) we find
\[
L_k \rho^{1/2} = \sum_{j}\theta \bar{u}_{kj}\rho^{1/2}L_j^* \theta 
               = \sum_{j} {u}_{kj}\rho^{1/2}\theta L_j^* \theta 
\]

Writing $\rho^{1/2}\theta L_j^* \theta$ as 
$\sum_m u_{jm} {L_m}\rho^{1/2}$ by condition \ref{tr-sdb-2} 
we have then
\[
L_k \rho^{1/2} = \sum_{j,m} {u}_{kj}u_{jm}L_m\rho^{1/2}
= \sum_m (u^2)_{km}L_m\rho^{1/2}
\]
which implies that $u^2=\unit\,$ by the linear independence of 
the $L_m\rho^{1/2}$. Therefore, 
since $u$ is unitary, $u=u^*$.

Conversely, if $\ref{tr-sdb-1}$ and $\ref{tr-sdb-2}$ 
hold, we can write 
$\rho^{1/2}\theta\Ll(\theta x\theta)\theta\rho^{1/2}$ as  
\begin{eqnarray*}
& & \rho^{1/2}\theta G^*\theta x \rho^{1/2} 
   + \sum_k \rho^{1/2} \theta L^*_k\theta x\theta L_k\theta \rho^{1/2}
   + \rho^{1/2} x \theta G\theta\rho^{1/2} \\
& = & G \rho^{1/2} x \rho^{1/2} 
   + \sum_j L_j \rho^{1/2} x \rho^{1/2} L_j^* 
   + \rho^{1/2} x \rho^{1/2} G^*.
\end{eqnarray*}
This, by Theorem \ref{th-dualesimm}, can be written as 
\[
\rho^{1/2}(G^\prime)^* x \rho^{1/2} 
+\sum_j \rho^{1/2}(L^\prime_j)^* x L^\prime_j \rho^{1/2}
+\rho^{1/2}x G^\prime \rho^{1/2}
= \rho^{1/2} \Ll^\prime(x)\rho^{1/2}
\]
It follows that $\theta\Ll(\theta x\theta)\theta
= \Ll^\prime(x)$ for all $x\in\Bh$ because $\rho$ is faithful. 
Moreover, it is easy to check by induction that 
$\theta\Ll^n(\theta x\theta)\theta = (\Ll^\prime)^n(x)$ for all $n\ge 0$. 
Therefore $\theta\T_t(\theta x\theta)\theta = \T^\prime_t(x)$ for 
all $t\ge 0$ and $\T$ satisfies the SQDB-$\theta$ condition by 
Proposition \ref{prop-TR-SDBsstL}. \qed 

\medskip

We now provide a geometrical characterisation of the SQDB-$\theta$
condition as in Theorem \ref{th-SQDB-geo}. To this end we introduce 
the trace class operator $R$ on $\mathsf{k}$ 
\begin{equation}\label{eq-operatot-R}
R_{jk}=\tr\left( \rho^{1/2}L^*_j \rho^{1/2} \theta L^*_k\theta\right)
\end{equation}
A direct application of Lemma \ref{lem-2bases} shows that $R$ is 
trace class. Moreover it is self-adjoint because, by the 
property $\tr(\theta x\theta)=\tr(x^*)$ of the antilinear time reversal, 
we have 
\begin{eqnarray*}
\overline{R}_{jk} 
& = & \overline{\tr\left( \rho^{1/2}L^*_j \rho^{1/2} \theta L^*_k\theta\right)} \\
& = & \tr\left( \theta (L_k\theta\rho^{1/2}L_j \rho^{1/2}\theta) \theta \right)\\
& = & \tr\left( \rho^{1/2}\theta L_j^*\rho^{1/2}\theta L^*_k\right) \\
& = & \tr\left( (\rho^{1/2}\theta L_j^*\theta)(\rho^{1/2} L^*_k)\right) = R_{kj}.
\end{eqnarray*}

\begin{theorem}\label{th-SQDB-theta-geo}
$\T$ satisfies the SQDB-$\theta$ if and only if the operators $G,L_k$ 
of a special GKSL of the generator $\Ll$ fulfill the following 
conditions:
\begin{enumerate}
\item\label{cnd-SQDBtheta-i} $\rho^{1/2}\theta{G^*}\theta
           =G\rho^{1/2}$,
\item\label{cnd-SQDBtheta-ii} the closed linear span of 
$\left\{\rho^{1/2}\theta L_\ell^*\theta\,\mid\,\ell\ge 1\right\}$ 
and $\left\{L_\ell \rho^{1/2}\,\mid\,\ell\ge 1\right\}$ in the 
Hilbert space of Hilbert-Schmidt operators on $\mathsf{h}$ coincide,
\item\label{cnd-SQDBtheta-iii} the self-adjoint trace class operators 
$R,C$ defined by (\ref{eq-ops-B&C-on-k}) and (\ref{eq-operatot-R}) 
commute and $C^{-1}R$ is unitary and self-adjoint.
\end{enumerate}
\end{theorem}
\noindent{\bf Proof.} It suffices to show that conditions 
\ref{cnd-SQDBtheta-ii} and \ref{cnd-SQDBtheta-iii} above are 
equivalent to condition \ref{tr-sdb-2} of Theorem \ref{th-SQDB-TR}. 

If $\T$ satisfies the SQBD-$\theta$, then it can be shown as in 
the proof of Theorem \ref{th-SQDB-geo} that \ref{cnd-SQDBtheta-ii} 
follows from condition \ref{tr-sdb-2} of Theorem \ref{th-SQDB-TR}.  
Moreover, left multiplying by $\rho^{1/2}L^*_\ell$ the identity  
$\rho^{1/2}\theta L^*_k\theta=\sum_{j}u_{kj}L_j\rho^{1/2}$ and 
tracing, we find 
\[
\tr\left(\rho^{1/2}L^*_\ell\rho^{1/2}\theta L^*_k\theta\right) 
= \sum_{j}u_{kj} \tr\left(\rho L^*_\ell L_j\right) 
\]
for all $k,\ell$ i.e. $R=CU^T$. The operator $U^T$ is also 
self-adjoint and unitary. Therefore $R$ and $C$ have the same range  
and, since the domain of $C^{-1}$ coincides with the range of $C$, 
the operator $C^{-1}R$ is everywhere defined, unitary and self-adjoint. 
It follows that the densely defined operator $RC^{-1}$ is a 
restriction of $(C^{-1}R)^*=C^{-1}R$ and $CR=RC$.  

In order to prove, conversely, that \ref{cnd-SQDBtheta-ii} and 
\ref{cnd-SQDBtheta-iii} imply condition \ref{tr-sdb-2} of 
Theorem \ref{th-SQDB-TR}, we first notice that, by the spectral 
theorem there exists a unitary $V=(v_{mn})_{m,n\ge 1}$ on the 
multiplicity space $\mathsf{k}$ such that $V^*CV$ is diagonal. 
Choosing a new GKSL representation of the generator $\Ll$ by 
means of the operators ${L''}_k = \sum_{n\ge 1}v_{nk}L_n$, if 
necessary, we can suppose that both $(L_\ell\rho^{1/2})_{\ell\ge 1}$ 
and $(\rho^{1/2} L_k^*)_{k\ge 1}$ are {\em orthogonal} bases of 
the same closed linear space. Note that 
\[
\tr\left(\rho^{1/2}(L'')^*_k\rho^{1/2}\theta(L'')^*_j\theta\right) 
=\sum_{m,n\ge 1}\bar{v}_{nk} v _{mj} 
\tr(\rho^{1/2}L^*_n\rho^{1/2}\theta L^*_m\theta)
\]
and the operator $R$, in the new GKSL representation, 
transforms into $V^*R V$ which is also self-adjoint.

Expanding $\rho^{1/2}\theta L_k^*\theta$ with respect to the 
orthogonal basis $(L_\ell\rho^{1/2})_{\ell\ge 1}$, 
for all $k\ge 1$, we have 
\begin{equation}\label{eq-expansionLk-theta}
\rho^{1/2}\theta L_k^*\theta = \sum_{\ell\ge 1}
\frac{\tr(\rho^{1/2}L^*_\ell\rho^{1/2}\theta L^*_k\theta)}
{\Vert L_\ell\rho^{1/2}\Vert_{HS}^2}\,L_\ell\,\rho^{1/2}
\end{equation}
i.e. $\rho^{1/2}L_k^*=\sum_{\ell} y_{k\ell} L_\ell\rho^{1/2}$ 
with a matrix $Y$ of complex numbers $y_{k\ell}$. 

Clearly, since $C$ is diagonal and commutes with $R$, we have  
$y_{k\ell}=(RC^{-1})_{k\ell}=(C^{-1} R)_{k\ell}$.
It follows then from \ref{cnd-SQDBtheta-iii} above that $Y$ 
coincides with the unitary operator $C^{-1}R$ and condition 
\ref{tr-sdb-2} of Theorem \ref{th-SQDB-TR} holds. Moreover, 
$Y$ is self-adjoint because both $R$ and $C$ are. \qed 

\medskip 
As an immediate consequence of the commutation of $R$ and $C$ we have 
the following parallel of Corollary \ref{cor-weightLl=weightLk} for 
the SQDB condition 

\begin{corollary}\label{cor-weightLl=weightLk-theta}
Suppose that a QMS $\T$ satisfies the SQDB-$\theta$ condition. 
For every special GKSL representation of $\Ll$ with operators 
$L_\ell\rho^{1/2}$ orthogonal as Hilbert-Schmidt operators on $\mathsf{h}$ 
if $\tr(\rho^{1/2}L^*_\ell\rho^{1/2}\theta L^*_k\theta)\not=0$  
for a pair of indices $k,\ell\ge 1$, then
$\tr(\rho L_\ell^*L_\ell)=\tr(\rho L_k^*L_k)$.
\end{corollary}

% \noindent{\bf Proof.} This follows immediately from self-adjointness 
% of the above $Y=C^{-1}R$ in (\ref{eq-expansionLk-theta}). Indeed, 
% $y_{k\ell}=y_{\ell k}$ together with 
% \[
% \overline{\tr\left(\rho^{1/2}L^*_k\rho^{1/2}\theta L^*_\ell\theta \right)}
% =\tr\left(\theta\left(L_\ell\rho^{1/2}\theta 
% L_k\theta\rho^{1/2}\right)\theta \right)
% =\tr\left(\rho^{1/2}L^*_\ell\rho^{1/2}\theta L^*_k\theta \right)
% \]
% because $\overline{\tr(x)}=\tr(x^*)=\tr(\theta x \theta)$. \qed
% {\tt NOTA}
% $ \rho^{1/2}L_\ell^* \to L_\ell\rho^{1/2} $
% {\tt antiunitary}

When the time reversal $\theta$ is given by the conjugation 
$\theta u = \bar{u}$ (with respect to some orthonormal basis 
of $\mathsf{h}$), $\theta x^*\theta$ is equal to the transpose 
$x^T$ of $x$ and we find the following 

\begin{corollary}\label{cor-SQDB-T} 
${\mathcal{T}}$ satisfies the {SQDB-$\theta$} condition 
if and only if there exists a special GKSL representation of 
$\Ll$, with operators $G, L_k$, such that:
\begin{enumerate}
\item \label{coniugio-1} $\rho^{1/2}G^T=G\rho^{1/2}$;
\item \label{coniugio-2} $\rho^{1/2}{L}_k^T = \sum_j u_{kj}L_j\rho^{1/2}$ 
for some unitary self-adjoint $(u_{kj})_{kj}$.
\end{enumerate}
\end{corollary}

\section{SQDB-$\theta$ for QMS on $M_2(\mathbb{C})$}

In this section, as an application, we find a standard form of a special 
GKSL representation of the generator $\Ll$ of a QMS on $M_2(\mathbb{C})$ 
satisfying the SQDB-$\theta$. 

\noindent The faithful invariant state $\rho$, in a suitable basis, 
can be written in the  form 
\[
\rho=\left(\begin{array}{cc}
            \nu & 0  \\   0 & 1-\nu
           \end{array}
     \right)
=\frac{1}{2}\left(\sigma_0+(2\nu-1)\sigma_3\right),\quad\quad 0<\nu<1
\] 
where $\sigma_0$ is the identity matrix and $\sigma_1,\sigma_2,\sigma_3$ 
are the Pauli matrices 
\[
\sigma_1=\left(\begin{array}{cc}
            0 & 1  \\   1 & 0
           \end{array}
     \right), \qquad
\sigma_2=\left(\begin{array}{cc}
            0 & -i  \\   i & 0
           \end{array}
     \right), \qquad
\sigma_3=\left(\begin{array}{cc}
            1 & 0  \\   0 & -1
           \end{array}
     \right).
\]
The time reversal $\theta$ is the usual conjugation in the above basis. 

In order to determine the structure of the operators $G$ and $L_k$ satisfying 
conditions of Corollary \ref{cor-SQDB-T} we find first a convenient basis of 
$M_2(\mathbb{C})$. We choose then a basis of eigenvectors 
of the linear map $X\to\rho^{1/2}X^T\rho^{-1/2}$ in $M_2(\mathbb{C})$ given by 
$\sigma_0,\sigma^{\nu}_1,\sigma^{\nu}_2, \sigma_3$ where 
\[
\sigma^{\nu}_1=\left(\begin{array}{cc}
            0 & \sqrt{2\nu}  \\   \sqrt{2(1-\nu)} & 0
           \end{array}
     \right), \qquad
\sigma^{\nu}_2=\left(\begin{array}{cc}
            0 & -i\sqrt{2\nu}  \\   i\sqrt{2(1-\nu)} & 0
           \end{array}
     \right).
\]
Indeed, $\sigma_0,\sigma^{\nu}_1, \sigma_3$ (resp. $\sigma^{\nu}_2$) are eigenvectors 
of the eigenvalue $1$ (resp. $-1$).  

Every special GKSL representation of $\Ll$ is given by (see \cite{FFVU} Lemma $6.1$)
\[
L_k=-(2\nu-1)z_{k3}\sigma_0+z_{k1}\sigma^{\nu}_1+z_{k2}\sigma^{\nu}_2+z_{k3}\sigma_3,
\quad\quad k\in\mathcal{J}\subseteq\{1,2,3\}
\]
with vectors $z_k:=(z_{k1},z_{k2},z_{k3})$ ($k\in\mathcal{J}$) linearly independent 
in $\mathbb{C}^3$.

The SQDB-$\theta$ holds if and only if $G,L_k$ satisfy
\begin{enumerate} 
\item[$(i)$] $G=\rho^{1/2}{G^T}\rho^{-1/2}$,
\item[$(ii)$] ${L}_k = \sum_{j\in\mathcal{J}} u_{kj}\rho^{1/2}{L_j^T}\rho^{-1/2}$ 
for some unitary self-adjoint $U=(u_{kj})_{k,j\in\mathcal{J}}$.
\end{enumerate}
Now, if $\mathcal{J}\not=\emptyset$, since every unitary self-adjoint matrix 
is diagonalizable and its spectrum is contained in $\{-1,1\}$, it follows 
that $U=W^*DW$ for some unitary matrix $W=(w_{ij})_{i,j\in\mathcal{J}}$ 
and some diagonal matrix $D$ of the form
\begin{equation}\label{formaD}
{\rm{diag}}(\epsilon_1,\ldots,\epsilon_{|\mathcal{J}|}),\quad
\quad\epsilon_i\in\{-1,1\},
\end{equation}
where $|\mathcal{J}|$ denotes the cardinality of $\mathcal{J}$.
Therefore, replacing the $L_k$'s by operators $L^\prime_k:=\sum_{j\in\mathcal{J}}w_{kj}L_j$ 
if necessary, we can take $U$ of the form \eqref{formaD}.

We now analyze the structure of $L_k$'s corresponding to the 
different (diagonal) forms of $U$. By condition $(ii)$ we have either 
$L_k=\rho^{1/2}{L_k^T}\rho^{-1/2}$ or $L_k=-\rho^{1/2}{L_k^T}\rho^{-1/2}$;
an easy calculation shows that 
\begin{equation}\label{1}
 L_k=\rho^{1/2}{L_k^T}\rho^{-1/2}
\qquad\hbox{if and only if}\qquad z_{k2}=0
\end{equation}
and
\begin{equation}\label{-1}
 L_k=-\rho^{1/2}{L_k^T}\rho^{-1/2} \quad\hbox{if and only if}\quad z_{k1}=z_{k3}=0.
\end{equation}
Therefore, the linear independence of $\{z_j:j\in\mathcal{J}\}$ forces 
$U$ to have at most two eigenvalues equal to $1$ and at most one equal 
to $-1$ and, with a suitable choice of a phase factor for each $L_k$, 
we can write
\begin{eqnarray}
L_k = (1-2\nu)r_k\sigma_0 + r_k\sigma_3 + \zeta_{k}\sigma_1^\nu
& \hbox{for} &  k=1,2 \hbox{ and  } r_k\in{\mathbb{R}}, 
\zeta_k\in{\mathbb{C}} \label{eq-L1-L2}   \\ 
L_3 = r_3 \sigma_2^\nu\hskip 1.64truecm
&  & \hbox{} \hskip 0.2truecm r_3\in{\mathbb{R}}. \label{eq-L3} 
\end{eqnarray}
Clearly $L_1$ and $L_2$ are linearly independent if and only if 
$r_{1}\zeta_{2}\not=r_{2}\zeta_{1}$. This, together with non triviality 
conditions leaves us, up to a change of indices, with the following possibilities:
\begin{itemize}
\item[$a)$] $|\mathcal{J}|=1$, $U=1$ then $\mathcal{J}=\{1\}$ with $r_1\zeta_1\not=0$,
\item[$b)$] $|\mathcal{J}|=1$, $U=-1$ then $\mathcal{J}=\{3\}$ with $r_3\not=0$,
\item[$c)$] $|\mathcal{J}|=2$, $U=\rm{diag}(1,1)$ then $\mathcal{J}=\{1,2\}$ 
with $r_1\zeta_1r_2\zeta_2\not=0$, $r_{1}\zeta_{2}\not=r_{2}\zeta_{1}$,
\item[$d)$] $|\mathcal{J}|=2$, $U=\rm{diag}(1,-1)$ then $\mathcal{J}=\{1,3\}$, 
with $r_3\not=0$, $r_1\zeta_1\not=0$, 
\item[$e)$] $|\mathcal{J}|=3$, $U=\rm{diag}(1,1,-1)$ then $\mathcal{J}=\{1,2, 3\}$ 
with $r_{1}\zeta_{2}\not=r_{2}\zeta_{1}$, $r_3\not=0$, $r_1\zeta_1r_2\zeta_2\not=0$.
\end{itemize}\smallskip

To conclude, we analyze condition $(i)$. If $G=\left(g_{jk}\right)_{1\le j,k\le 2}$
then statement $(i)$ is equivalent to 
\begin{equation}\label{eq-condG}
\sqrt{\nu}\,g_{21}=\sqrt{1-\nu}\,g_{12}.
\end{equation}
Since $G=-iH-2^{-1}\sum_kL^*_kL_k$ with $H=\sum_{j=1}^3v_j\sigma_j$, $v_j\in\mathbb{R}$, 
and $\sum_{k} L^*_kL_k$ is equal to the sum of a term depending only on
$\sigma_0$ and $\sigma_3$ plus
\[
\sum_{k=1,2}2 r_k \left(\begin{array}{cc}
0  &  \zeta_k\sqrt{2\nu}(1-\nu)- \bar{\zeta}_k\nu \sqrt{2(1-\nu)} \\
\bar{\zeta}_k\sqrt{2\nu}(1-\nu)- \zeta_k\nu \sqrt{2(1-\nu)} &  0\\
\end{array}\right)
\]
in the case $\mathcal{J}\not=\emptyset$ the identity (\ref{eq-condG}) holds if and only if 
\begin{equation}\label{eq-v1-v2}
\left\{\begin{array}{ccl}
v_1\left(\radm\right)&=&-\sqrt{2\nu(1-\nu)}
\left(\radp\right)^2\sum_{k=1}^2r_k\im\zeta_k\\
v_2\left(\radp\right)&=&-\sqrt{2\nu(1-\nu)} \left(\radm\right)^2 
\sum_{k=1}^2r_k\re\zeta_k
\end{array}\right..
\end{equation}
On the other hand, when $\mathcal{J}=\emptyset$, condition (\ref{eq-condG}) is equivalent to 
$\sqrt{\nu}(v_1+iv_2)=\sqrt{1-\nu}(v_1-iv_2)$, i.e.
\begin{equation}\label{eq-v1-v2-soloH}
v_1\left(\radm\right)=0, \qquad v_2=0
\end{equation}
Therefore we have the following possible standard forms for $\Ll$.

\begin{theorem}\label{th-strct-SQDB-theta-M2C}
Let $L_1, L_2, L_3$ be as in (\ref{eq-L1-L2}), (\ref{eq-L3}), 
$H=\sum_{j=1}^3v_j\sigma_j$ with $v_1, v_2$ as in (\ref{eq-v1-v2}) and $v_3\in{\mathbb{R}}$. 
The QMS $\T$ satisfies the SQDB-$\theta$ if and only if there exists a 
special GKSL representation of $\Ll$ given, up to phase factors multiplying 
$L_1, L_2, L_3$, in one of the following ways:
\begin{itemize}
\item[$o)$]  $H$ with $v_1=v_2=0$ if $\nu\not=1/2$, and $v_1\in\mathbb{R}$, $v_2=0$ if $\nu=1/2$,
\item[$a)$]  $H, L_1$ with $r_1\zeta_1\not=0$,
\item[$b)$]  $H, L_3$ with $r_3\not=0$,
\item[$c)$]  $H, L_1,L_2$ with $r_1\zeta_1r_2\zeta_2\not=0$ and $r_{1}\zeta_{2}\not=r_{2}\zeta_{1}$,  
\item[$d)$]  $H, L_1, L_3$ with $r_3\not=0$ and $r_1\zeta_1\not=0$,
\item[$e)$]  $H, L_1, L_2, L_3$ with $r_{1}\zeta_{2}\not=r_{2}\zeta_{1}$,  
$r_1\zeta_1r_2\zeta_2\not=0$ and $r_3\not=0$.
\end{itemize}
\end{theorem} 

Roughly speaking, the standard form of $\Ll$ corresponds, up to degeneracies when 
some of the parameter vanish or when some linear dependence arises, to the case e).

We know that a QMS satisfying the usual (i.e. with pre-scalar product with $s=0$) 
QDB-$\theta$ condition must commute with the modular group. Moreover, when 
this happens, the SQDB-$\theta$ and QDB-$\theta$ conditions are equivalent 
(see e.g. \cite{Cip}, \cite{FFVU}). 

We finally show how the generators of a QMSs on $M_2({\mathbb{C}})$ 
satisfying the usual QDB-$\theta$ condition can be recovered by a special choice 
of the parameters $r_1,r_2,r_3,\zeta_1,\zeta_2$ in Theorem \ref{th-strct-SQDB-theta-M2C} 
describing the generator of a QMS satisfying the SQDB-$\theta$ condition.

To this end, we recall that $\T$ fulfills the QDB-$\theta$ when 
$\tr(\rho x \T_t(y))=\tr(\rho\theta y^* \theta \T_t(\theta x^* \theta))$ 
for all $x,y\in\Bh$. In \cite{FFVU} we classified generators of QMS on 
$M_2({\mathbb{C}})$ satisfying the QDB condition without time reversal 
(i.e., formally, replacing $\theta$ by the identity operator, that is, 
of course, not antiunitary). The same type of arguments show that,  
disregarding trivialisations that may occur when some of the parameters 
below vanish, QMSs on $M_2({\mathbb{C}})$ satisfying the QDB-$\theta$ 
condition have the following standard form 
\begin{eqnarray}\label{eq-QMS-2x2-QDB-theta-standard-form}
& & \Ll(x) = i[H,x] - \frac{\mid\!\eta\!\mid^2}{2}\left(L^2x-2LxL+xL^2\right) \\
& & -\frac{\mid\!\lambda\!\mid^2}{2}\left(\sigma^-\kern-2truept\sigma^+x
      -2\sigma^-x\sigma^+ +x\sigma^-\kern-2truept\sigma^+\right)
-\frac{\mid\!\mu\!\mid^2}{2}\left(\sigma^+\sigma^-x-2\sigma^+x\sigma^- 
+x\sigma^+\sigma^-\right)\nonumber,
\end{eqnarray}
where $H=h_0\sigma_0+h_3\sigma_3$ ($h_0,h_3\in\mathbb{R}$), 
$L=-(2\nu-1)\sigma_0+\sigma_3$, $\sigma^\pm=(\sigma_1\pm i\sigma_2)/2$ and, 
changing phases if necessary, $\lambda,\mu,\eta$ can be chosen as {\em 
non-negative real} numbers satisfying 
\begin{equation}\label{eq-lambda-mu}
\lambda^2(1-\nu)=\nu {\mu^2}.
\end{equation}

Choosing $r_1=\eta, \zeta_1=0$ we find immediately that the operator 
$L$ in (\ref{eq-QMS-2x2-QDB-theta-standard-form}) coincides with the 
operator $L_1$ in (\ref{eq-L1-L2}). Moreover, choosing $r_2=0$ we find  
$v_2=0$ and also $v_1=0$ for $\nu\not=1/2$. 
A straightforward computation yields 
\[
\left(\begin{array}{cc}
            \lambda\, \sigma_{+} \\
            \mu\, \sigma_{-}\\
           \end{array}
          \right)=
 \left(\begin{array}{cc}
            \lambda/(2\zeta_2\sqrt{2\nu}) & i\lambda/(2r_3\sqrt{2\nu})\\
            \mu/(2\zeta_2\sqrt{2(1-\nu)}) & -i\mu/(2r_3\sqrt{2(1-\nu)}) \\
\end{array}
          \right)\left(\begin{array}{cc}
            {L}_2\\
            {L}_3\\
           \end{array}
          \right)
\]
and the above $2\times 2$ matrix is unitary if we choose $\zeta_2=\lambda/(2\sqrt{\nu})$, 
$r_3=i\mu/(2\sqrt{1-\nu)})=i\zeta_2$ because of (\ref{eq-lambda-mu}) and changing the phase 
of $r_3$ in order to find a unitary that is also self-adjoint.

This shows that we can recover the standard form (\ref{eq-QMS-2x2-QDB-theta-standard-form}) 
choosing $H$, $L_1,L_2,L_3$ as in Theorem \ref{th-strct-SQDB-theta-M2C} e) with 
$r_1=\eta, \zeta_1=0, r_2=0, \zeta_2=\lambda/(2\sqrt{\nu}), r_3=i\mu/(2\sqrt{1-\nu)}), 
v_1=v_2=0$.

\section*{Appendix}

We denote by $\ell^2(J)$ denote the Hilbert space of complex-valued, 
square summable sequences indexed by a finite or countable set $J$.

\begin{lemma}\label{lem-2bases}
Let $\mathcal{J}$ be a complex separable Hilbert space and let 
$(\xi_j)_{j\in J}$, $(\eta_j)_{j\in J}$ be two Hilbertian bases 
of $\mathcal{J}$ satisfying
$\sum_{j\in J} \left\Vert \xi_j\right\Vert^2 <\infty$, 
$\sum_{j\in J} \left\Vert \eta_j\right\Vert^2 <\infty$.
The complex matrices $A=(a_{jk})_{j,k\in J}$, $B=(b_{jk})_{j,k\in J}$, 
$C=(c_{jk})_{j,k\in J}$ given by 
\[
a_{jk}=\langle\xi_j,\xi_k\rangle, \quad
b_{jk}=\langle\xi_j,\eta_k\rangle, \quad 
c_{jk}=\langle\eta_j,\eta_k\rangle
\]
define trace class operators on $\ell^2(J)$ satisfying 
$B^*A^{-1}B = C$. Moreover $A$ and $C$ are self-adjoint 
and positive.
\end{lemma}

\noindent{\bf Proof.} Note that 
\[
\sum_{j,k\ge 1}\left|b_{jk}\right|^2 
\le \sum_{j,k\ge 1}
\left\Vert\xi_j\right\Vert^2\cdot\left\Vert\eta_k\right\Vert^2
= \sum_j\left\Vert\xi_j\right\Vert^2 \cdot
\sum_k \left\Vert\eta_k\right\Vert^2 <\infty
\]
Therefore $B$ defines a Hilbert-Schmidt operator on $\ell^2(J)$. 

In a similar way $A$ and $C$ define Hilbert-Schmidt operators on 
$\ell^2(J)$ that are obviously self-adjoint. These are also positive 
because for any sequence $(z_m)_{m\in J}$ of complex numbers with 
$z_m\not=0$ for a finite number of indices $m$ at most we have 
\[
\sum_{m,n\in J} \bar{z}_m a_{mn} z_n 
=\sum_{m,n\in J} \bar{z}_m \left\langle \xi_m,\xi_n \right\rangle z_n  
=\left\Vert \sum_{m\in J} z_m\xi_m\right\Vert^2 \ge 0.
\]
Moreover, they are trace class because 
\[
\sum_{j\in J} a_{jj} = \sum_{j\in J}\left\Vert \xi_j\right\Vert^2 < \infty, 
\qquad
\sum_{j\in J} c_{jj} = \sum_{j\in J}\left\Vert \eta_j\right\Vert^2 < \infty. 
\]
Finally, we show that $B$ is also trace class. By the spectral theorem, 
we can find a unitary $V=(v_{kj})_{k,j\in J}$ on $\ell^2(J)$ such that 
$V^*AV$ is diagonal. The series $\sum_{m\in J}v_{mj}\xi_m $ is  
norm convergent because 
\[
\left\Vert\sum_{m}v_{mj}\xi_m\right\Vert^2 
= \sum_{m,n\in J} \bar{v}_{nj}a_{nm}v_{mj} = (V^*AV)_{jj}. 
\]
The series $\sum_{m\in J}v_{mj}\xi_{m}$ is norm convergent as well 
for a similar reason. Therefore, 
putting $\xi^\prime_j = \sum_{m\in J} v_{mj}\xi_m$ and 
$\eta^\prime_j = \sum_{m\in J} v_{mj}\eta_m$ we find immediately 
$(V^*AV)_{kj}=\langle\xi^\prime_k,\xi^\prime_j\rangle=0$ for $j\not=k$,  
$(V^*AV)_{jj} = \left\Vert\xi^\prime_j\right\Vert^2$ and
\begin{eqnarray*}
(V^*BV)_{kj} 
& = & \sum_{m,n} \bar{v}_{mk}v_{nj}\langle \xi_m,\eta_j\rangle 
=\langle \xi^\prime_k,\eta^\prime_j\rangle,\\
(V^*CV)_{kj} 
& = &\sum_{m,n} \bar{v}_{mk}v_{nj}\langle \eta_m,\eta_j\rangle 
=\langle \eta^\prime_k,\eta^\prime_j\rangle.
\end{eqnarray*}
As a consequence, the following identity
\begin{eqnarray*}
\left(V^*B^*A^{-1}BV\right)_{kj}
& = &\left((V^*B^*V)(V^*AV)^{-1}(V^*BV)\right)_{kj} \\
& = & \sum_{m\in J}(V^*B^*V)_{km}\left((V^*AV)_{mm}\right)^{-1}(V^*BV)_{mj} \\
& = & \sum_{m\in J}\left\langle {\eta^\prime_k},
   \frac{\xi^\prime_m}{\Vert\xi^\prime_m\Vert}\right\rangle 
    \left\langle \frac{\xi^\prime_m}{\Vert\xi^\prime_m\Vert}, 
    \eta^\prime_j\right\rangle \\
& = &  \langle \eta^\prime_k,\eta^\prime_j\rangle = (V^*CV)_{kj}
\end{eqnarray*}
holds because $(\xi^\prime_m/\Vert\xi^\prime_m\Vert)_{m\in J}$ 
is an orthonormal basis of ${\mathcal{J}}$. 

This proves that $V^*B^*A^{-1}BV=V^*CV$ i.e. $B^*A^{-1}B=C$. 
It follows that $|A^{-1/2}B|=C^{1/2}$ is Hilbert-Schmidt as 
well as $A^{-1/2}B$ and $B = A^{1/2}(A^{-1/2}B)$ is trace class 
being the product of two Hilbert-Schmidt operators. \qed

\bigskip
\noindent{\bf Acknowledgment.} The financial support form the 
MIUR PRIN 2007 project ``Quantum Probability and Applications 
to Information Theory'' is gratefully acknowledged.

\end{document}